\documentclass[floats,amsmath,amssymb,prb,aps,twocolumn]{revtex4}
\usepackage{graphics,amsmath,graphicx}
\usepackage{txfonts}
\usepackage[colorlinks=true,citecolor=blue,linkcolor=blue,urlcolor=blue]{hyperref}

\begin{document}
\title{On dynamical localization corrections to band transport}

\author{S. Fratini$^{1}$, S. Ciuchi$^{2,3}$}

\affiliation{$^1$Universit\'e Grenoble Alpes, CNRS, Grenoble INP, Institut N\'eel, 38000 Grenoble, France \\
$^2$Dipartimento di Scienze Fisiche e Chimiche,
Universit\`a dell'Aquila, via Vetoio, I-67010 Coppito-L'Aquila, Italy\\
$^3$ Istituto dei Sistemi Complessi - CNR, Via dei Taurini 19, I-00185 Rome, Italy}

\begin{abstract}
Bloch-Boltzmann transport theory fails to describe the carrier diffusion
in current crystalline organic semiconductors, 
where the presence of large-amplitude thermal molecular motions 
causes substantial dynamical disorder.
The charge transport mechanism in this original situation
is now understood in terms of a 
transient localization of the carriers' wavefunctions, whose applicability is however limited to the
strong disorder regime. 
In order to deal with the ever-improving 
performances of new materials,
we develop here 
a unified theoretical framework that 
includes transient localization theory as a limiting case, and smoothly connects with 
the standard band description when
molecular disorder is weak.
The theory, which specifically adresses the emergence
 of dynamical localization corrections to semiclassical transport,
is used to determine a "transport phase diagram" of 
high-mobility organic semiconductors.
\end{abstract}

\date{\today}
\maketitle
\section{Introduction}

The last decade has witnessed considerable progress in the understanding of
charge transport in high-mobility organic semiconductors, with
important milestones
achieved in both experimental and theoretical research.
On the experimental side,
the widespread access and improved control on field-effect devices has provided a common ground
for the systematic and reproducible measurement of carrier mobilities \cite{Choi}.
Initially restricted to crystalline rubrene \cite{Podzorov,Frisbie}, which served as
a prototypical material
due to its outstanding performances and stability,
results indicative of intrinsic charge transport
are now  obtained in a
growing number of organic semiconductors
\cite{Minder,OkamotoT,Hofmockel,Mitsui,Krupskaya,Niazi,Illig,Kubo}.
On the theoretical side, it is now understood that 
the dominant intrinsic factor limiting the  mobility 
is the presence of large thermal vibrations of the constituent molecules, which cause strong dynamic disorder 
\cite{Troisi06}. 
The latter hinders the carrier motion in ways that
differ substantially from what predicted by semiclassical scattering theories.

The dynamical nature of molecular disorder in organic solids makes
the quantum theory of Anderson localization, that was developed for systems with static randomness 
\cite{RMP84,MottKaveh}, of 
limited use \textit{per se}.
To deal with this original situation, the physical idea of charge carriers being
coherently localized, but only over a limited timeframe, has emerged over the 
years \cite{Troisi06,Ciuchi11,Fratini-AFM16}. By highlighting explicitly a
connection with the physics of localized systems, the concept of
\textit{transient localization} could reconcile a number of puzzling features of
organic semiconductors, most notably the observation of "band-like" mobilities
decreasing with temperature to values below the so-called Mott-Ioffe-Regel
limit. The applicability of the transient localization approach, however, is by construction
restricted to the regime of strong dynamical disorder. Materials with reduced
localization effects, either featuring more isotropic two-dimensional band structures \cite{NMat17}
or lower degrees of disorder \cite{Alkylation}, are actively
investigated. It can be expected that future organic compounds will progressively 
move away from the strong disorder regime, entering a crossover region
for which there is yet no available theoretical description.

The aim of this work is to establish a unified
theoretical framework that encompasses the whole range from Bloch-Boltzmann band
theory, that applies in the limit of weak electron-phonon scattering
\cite{Glarum,Friedman,Cheng,Bernardi}, to the transient localization (TL) regime
relevant when dynamic disorder is strong \cite{Fratini-AFM16}. 
Our approach, which is based on the evaluation of dynamical localization corrections  
to semiclassical transport, is valid regardless of the disorder strength, as confirmed by the comparison with 
available exact numerical data.
We illustrate our findings 
on organic compounds of current interest,  determining a general  "transport phase diagram" for 
high-mobility organic semiconductors.

\section{Theoretical methodology}

\subsection{Dynamical localization corrections}
\label{sec:dynloc}

Due to its semiclassical nature,  
Bloch-Boltzmann theory neglects localization processes altogether.
 While this is a viable 
approximation to treat electron-lattice interactions in the weak scattering limit,
it is inappropriate 
to study the transport properties of electrons in strongly fluctuating environments, 
as is the case in organic semiconductors owing to the presence of large-amplitude molecular motions.
Here we want to overcome this limitation
by restoring 
those quantum  processes that are missing in the semiclassical description.
Our derivation is based on Kubo response theory, and builds  on the formulation 
developed in Refs. \cite{Ciuchi11,Ciuchi12}.
The key quantity of interest is the time-dependent velocity-velocity
anticommutator correlation function, $C(t)=\langle \lbrace \hat{V}(t),\hat{V}(0)\rbrace\rangle$, with
$\hat{V}$ the velocity operator for charge carriers in a given direction. The dynamical observables 
describing charge transport, i.e. the charge diffusivity, the charge mobility and the optical conductivity,
can all be derived from the knowledge of this time-dependent quantity\cite{Ciuchi11,Ciuchi12}.

Let us imagine that for a given system of electrons 
interacting with lattice vibrations, we are able to calculate the velocity correlation function using  
semiclassical Bloch-Boltzmann theory (see Appendix \ref{sec:band}),  that we denote as $C_{SC}(t)$. 
By definition this quantity  misses all those quantum processes that are instead present in the exact
correlator $C(t)$. It is then natural to define as \textit{dynamical localization corrections} (DLC)
the difference $\delta C(t)=C(t)-C_{SC}(t)$  between the 
exact correlator and the semiclassical result. $\delta C(t)$ 
 obviously entails all those velocity correlations that are not included in the semiclassical description.
While this quantity
is generally unknown, as its calculation requires the full solution of the problem,
we now provide an approximation scheme that has a very broad validity and that turns out to be quantitatively accurate 
for the problem at hand, where the dynamical disorder is slow compared to the free-carrier dynamics.

The key observation is that the quantum corrections $\delta C$ induced by dynamical lattice fluctuations are  
continuously connected
to those that would be realized in a perfectly frozen lattice environment, $\delta C_0$: the latter should  be adiabatically  
recovered when the timescale of lattice fluctuations is sufficiently long as compared to all other timescales 
in the problem.  The similarity between $\delta C$ and $\delta C_0$ for slowly fluctuating disorder is advantageous, because 
solving a problem with frozen disorder (which can be done exactly via 
the diagonalization of a one-body disordered Hamiltonian) 
is a much easier task than solving the full dynamical problem, 
which is instead prohibitively difficult.
The quantity $\delta C_0=C_0-C_{SC}$ evaluated in the frozen disorder limit contains most of the information that 
we need on localization corrections. The small difference between $\delta C_0$ and $\delta C$, originating solely from the dynamical 
nature of the lattice, can then be treated in an approximate way. The power of the method relies on the fact that
no assumptions are made on the smallness of  
$\delta C$ itself, so that the whole approach is not restricted to the weak disorder limit.
 
\subsection{Decorrelation time}

Disorder dynamics (here, the dynamics of atomic and molecular positions) are known to destroy the quantum processes at the very
origin of wavefunction localization  \cite{RMP84,MottKaveh}. 
In the scaling theory of localization, the motion of the scattering centers on the dynamical 
timescale $\tau_{d}$
gives rise to a finite cutoff length 
\cite{Thouless} which corresponds to the 
length traveled by semiclassical particles over this time
\cite{RMP84};
this  cutoff length/time restores a finite diffusivity for the carriers, which would otherwise be vanishing.
When translated to our problem, this implies that the quantum corrections $\delta C_0$ characterizing 
the localized system can only be sustained at times that are short compared 
to the timescale of dynamic disorder \cite{Shante78,Gogolin87}, 
while they  decay and vanish at longer times (see Appendix \ref{sec:Thouless}).
This is embodied in the 
following form:
\begin{equation}
\delta C= \delta C_0 e^{-t/\tau_{d}},
\label{eq:RTE}
\end{equation}
with the decorrelation time being set by the frequency of the relevant modes, i.e. 
$\tau_{d}\sim 1/\omega_0$ within a numerical factor (the optimal value of such prefactor will be discussed below).
The corresponding velocity correlator is then
\begin{equation}
C(t)= C_{SC}(t)+ \delta C_0(t) e^{-t/\tau_{d}}.
\label{eq:CRTE}
\end{equation}
Eq. (\ref{eq:CRTE}) constitutes the basis of our theoretical approach, that will now be benchmarked and applied to the 
study of charge transport in organic materials in the next sections.

Before proceeding further, we argue that the proposed approximation scheme
should be valid even beyond the slow disorder limit initially assumed in the derivation. The reason is that
Eq. (\ref{eq:CRTE}) is able to 
interpolate from full localization all the way to the semiclassical limit depending on the value of $\tau_{d}$. 
Indeed, while  letting  $\tau_{d}\to \infty$  obviously restores the correlator of the localized system, 
$C(t)=C_0(t)$,   taking the opposite limit of fast disorder, $\tau_{d}\to 0$, 
suppresses all quantum corrections, hence leaving $C(t)=C_{SC}(t)$.
More generally, in cases where localization corrections are irrelevant to start with
(because lattice fluctuations are weak, i.e. for weak electron-phonon interactions and at low temperatures), then
$\delta C_0$ can be set to zero. 
Correspondingly, the correct semiclassical result will be trivially   recovered by Eq. (\ref{eq:CRTE})
regardless of the value of $\tau_{d}$. 

\subsection{Carrier mobility}

The carrier mobility 
can now be obtained from Eq. (\ref{eq:CRTE}) following the lines of Ref.  \onlinecite{Ciuchi11}. 
We first observe that the diffusion constant $\mathcal{D}$ is the long-time limit
of the instantaneous diffusivity, defined as $D(t)=\int_0^t C(t^\prime) dt^\prime/2$.
Introducing the Laplace transform $\tilde{C} (p)=\int_0^\infty C(t) e^{-pt} dt$, one has 
\begin{equation}
\mathcal{D}=\lim_{t\to\infty}D(t)=\tilde{C} (0)/2. 
\end{equation}
Applying the Laplace transform to Eq. (\ref{eq:CRTE})
yields 
\begin{equation}
\tilde{C}(0)=\tilde{C}_{SC}(0)+ \delta \tilde{C}_0(p),
\label{eq:Clap}
\end{equation}
with $p=1/\tau_{d}$, 
and hence $\mathcal{D}=\mathcal{D}_{SC}+\delta \tilde{C}_0(p)/2$.
The mobility is then obtained from Einstein's relation $\mu=e\mathcal{D}/k_BT$ as
\begin{equation}
\mu =  \mu_{band}+ \delta \mu,
\label{eq:muRTE}
\end{equation}
with $\mu_{band}=\frac{e\mathcal{D}_{SC}}{k_BT}$ and $\delta \mu=\frac{e}{2k_BT} \delta \tilde{C}_0(p)$. 
The above expression has the desired form of a semiclassical band mobility  
corrected by quantum processes.

To make the connection with previous works, we observe that 
when  the term $\delta \mu$ dominates in Eq. (\ref{eq:muRTE}), i.e. when
localization effects are sufficiently strong,  one is allowed 
to neglect the semiclassical terms altogether, which  corresponds to
setting $C =\delta C$ in the previous derivation.  Eq. (\ref{eq:CRTE}) then becomes 
$C(t)=   C_0(t) e^{-t/\tau_{d}}$, which is  the form that was originally assumed in Ref. \onlinecite{Ciuchi11},
setting the basis of transient localization theory.
Repeating the same steps as above,   Eq. (\ref{eq:muRTE}) reduces to 
the usual TL formula for the charge mobility, $\mu=e pL^2(p)/2k_BT$, with the transient localization length 
defined by $L^2(p)=\tilde{C}_0(p)/p$
\cite{Ciuchi11,Fratini-AFM16}.

For the practical implementation of Eq. (\ref{eq:muRTE}), 
we actually rearrange Eq. (\ref{eq:Clap}) as 
\begin{equation}
\tilde{C}(0)=\tilde{C}_{0}(p)+[\tilde{C}_{SC}(0) -\tilde{C}_{SC}(p)],
\label{eq:Clap2}
\end{equation}
which allows us to take full advantage of the 
numerical tools already developed in the context of transient localization theory. The first term in Eq. (\ref{eq:Clap2}) is then easily recognized as the TL result, which is readily evaluated with the methods described in Refs. \onlinecite{Nemati} and \onlinecite{NMat17}. The remaining terms between
brackets now only involve semiclassical quantities, which are evaluated following the procedure described in Appendix A. Finally, the mobility Eq. (\ref{eq:muRTE}), is obtained making use of the explicit formula Eq. (\ref{eq:mupractical}), which follows directly from Eq. (\ref{eq:Clap2}).
Full details are presented in Appendix \ref{sec:band}, \ref{sec:crossover} and \ref{sec:nummethods}.

\section{Results}

\subsection{Models}
The theory developed in the preceding sections is totally general and can be applied 
to a variety of electron-phonon interaction models. 
For illustrative purposes we shall mainly consider the following class of tight-binding Hamiltonians\cite{NMat17}. 
\begin{equation}
H=\sum_{i,\delta}  [J_\delta +\alpha_\delta x_{i,\delta}] (c^+_i c_{i+\delta} +h.c.) + H_x,
\label{eq:H}
\end{equation}
which is broadly representative of the  physics of organic semiconductors.
Eq. (\ref{eq:H}) describes charge carriers moving on a  molecular lattice, with 
nearest neighbor transfer integrals $J_\delta$ in the different bond directions  $\delta$ that are
linearly modulated by the coupling to intermolecular modes $x_{i,\delta}$.
Unless otherwise specified, the modes are assumed to be uncorrelated between different bonds, as described 
by the Hamiltonian $H_x=\sum_{i,\delta} K x_{i,\delta}^2/2+ p_{i,\delta}^2/2M$, where  $\omega_0=\sqrt{K/M}$
is the typical frequency of intermolecular vibration. 
Correlated bond fluctuations, as well as Holstein-type local interactions with slow intramolecular modes, 
of the form $H_I=\sum_i \alpha_H  c^+_i c_i x_i$,
can also be studied within the present general theoretical framework (see below and Appendix \ref{sec:models}). 
The interaction with high-frequency intra-molecular modes, which is not treated explicitly here, 
can be included  via a rescaling of the transfer integrals $J_\delta$, corresponding to the usual polaronic 
band narrowing \cite{Ren}.

Following Ref. \onlinecite{NMat17} we consider a two-dimensional, hexagonal molecular lattice of unit spacing $a$, 
with nearest neighbors $\delta=$a,b,c 
(see the sketch in Fig. \ref{fig:diffcomp}).
We take $J=\sqrt{J_a^2+J_b^2+J_c^2}$ as the energy unit, which fixes the scale of the band dispersion. 
For clarity of presentation, we shall focus on the 
common situation encountered in 
high-mobility molecular semiconductors,  
where two bond directions are equal by symmetry,
so that the set of transfer integrals can be characterized by
a single parameter $\theta$ ($J_a=J \cos{\theta}$, $J_b=J_c=J \sin{\theta}/\sqrt{2}$)  \cite{NMat17}.
Moreover, we shall be mostly concerned  
with the high-temperature regime where $k_BT\gtrsim \hbar\omega_0$. In this case the mean square 
thermal fluctuation of the transfer integrals is readily evaluated in terms of the parameters of Eq. (\ref{eq:H}) to  
$\Delta J_{\delta} = \alpha_\delta \langle x_{i,\delta}^2 \rangle^{1/2}= \alpha_\delta (k_BT/K)^{1/2}$. We can then
introduce $\Delta J = \sqrt{\Delta J_a^2+\Delta J_b^2+\Delta J_c^2}$ as a measure of the overall energetic disorder
induced by the intermolecular displacements. Unless otherwise specified, we fix the microscopic parameters to $J=0.1$eV 
and $T/J=0.25$, representative of high-mobility organic 
semiconductors at room temperature\cite{NMat17}, and set $\hbar=1$.

\subsection{Validation of the theory}

Before exploring our general findings we validate the theory 
by benchmarking it against  exact results available on simplified one-dimensional models. We consider
the quantum Monte Carlo (QMC) data of Ref. \onlinecite{DeFilippis-PRL15}, and  the Finite Temperature
Lanczos Method (FTLM) data of Ref. \onlinecite{Fehske-PRB05}.
Both approaches are in principle exact, and they fully include
the quantum dynamics of the molecular vibrations without any assumptions. 
The case studied in Ref. \onlinecite{DeFilippis-PRL15} corresponds to the one-dimensional 
limit of the model Eq. (\ref{eq:H}) in the presence of maximal
correlations between neighboring intermolecular motions 
(see e.g. Refs. \onlinecite{Troisi06,Fratini09,Ciuchi11,Nemati} and Appendix \ref{sec:models}), in a regime 
where localization corrections are weak. 
The results of Ref. \onlinecite{Fehske-PRB05} are instead for the one-dimensional Holstein model, where only on-site 
(intramolecular) interactions are considered (Appendix \ref{sec:models}), in a regime where localization corrections
are strong.

\begin{figure}
   \centering
   \includegraphics[width=8cm]{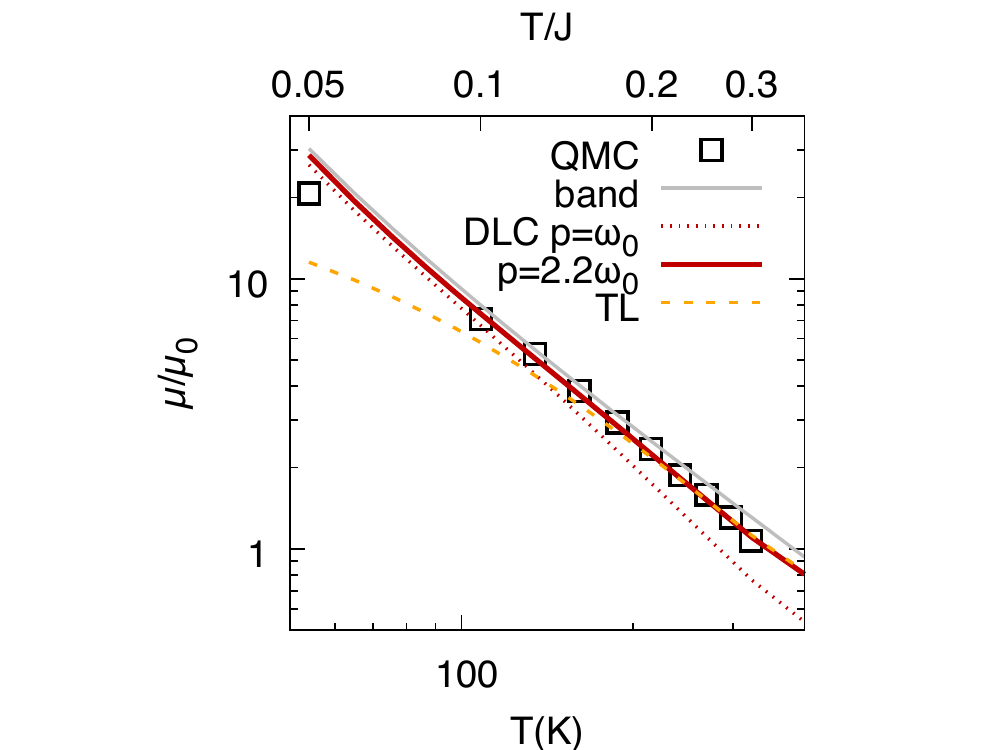}
   \includegraphics[width=7cm]{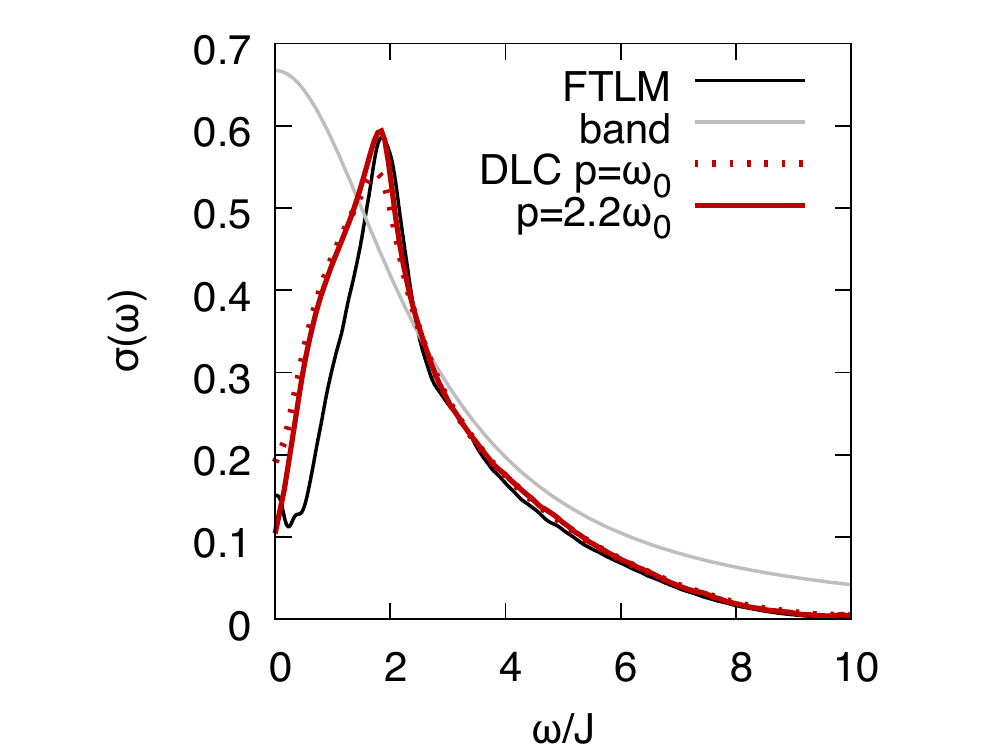}
 \caption{(a) Mobility calculated with dynamical localization corrections from Eq. (\ref{eq:muRTE}) (DLC, red)
 on a one-dimensional chain with correlated bond-disorder (see text). Squares are the QMC data of Ref. \onlinecite{DeFilippis-PRL15}.
 The band result derived in Appendix \ref{sec:band} [Eq. (\ref{eq:rateSSH})] and the TL result  are also shown (thin gray and orange dashed line respectively).
 The microscopic parameters are $J=93$meV, $\omega_0/J=0.05$ and $\lambda=0.17$, as defined in Appendix \ref{sec:models}. (b) Optical conductivity 
calculated for a one-dimensional chain with
 local intramolecular interactions (red), compared with the FTLM data of Ref. \onlinecite{Fehske-PRB05} (black) 
 and the band result [thin gray line, from Eq. (\ref{eq:rateHol})]. Parameters are $\omega_0/J=0.2$, $\lambda=1$ and $T/J=0.5$.} 
\label{fig:validation}
 \end{figure}

Fig.  \ref{fig:validation}(a) shows the QMC data for the mobility versus temperature (squares) 
together with the result obtained from the present theory
(DLC, red full and dotted lines) 
and band theory, as described in Appendix \ref{sec:band}
 [Eq. (\ref{eq:rateSSH})] (gray, thin). 
An excellent quantitative agreement with the 
QMC data is obtained if one takes a value $p=1/\tau_d=2.2\omega_0$ in the DLC theory (red, full line). To illustrate the impact
of the decorrelation time on the mobility, we also show the result obtained for
$p=\omega_0$ (red dotted). Reducing the value of $p$ tends to 
overestimate quantum localization effects, and hence to underestimate
the mobility. 
We observe that  with the present choice of model parameters, which corresponds to a moderate amount of molecular disorder
($\Delta J/J= 0.41$ at $T/J=0.25$),
the QMC result is itself qualitatively similar to the band theory result in the whole temperature range explored, and the
reduction of the mobility by localization corrections is less than $15\%$. 
rubrene 
We also note that the QMC data  do not recover exactly the calculated band value in the low temperature limit, 
where molecular fluctuations are suppressed.
This could signal either the presence of scattering processes not included in the perturbative 
band result, or a numerical artefact brought by the analytical continuation in the QMC calculation.  Finally, 
the TL result (orange dashed) is quantitatively accurate around room temperature, but it becomes inappropriate in the 
weak disorder regime attained at lower temperatures (see next Sections).

Fig.  \ref{fig:validation}(b) shows the exact FTLM data for the optical conductivity per particle 
 (black, thin),
together with the result obtained from the present theory (DLC, red full and dotted lines) and 
from band theory (gray, thin) (details on the calculation of the frequency-dependent response are provided
in Appendix \ref{sec:crossover}).
Because in this example the level of disorder is quite large (the thermal fluctuation of the 
on-site molecular energy is evaluated to $\Delta/J=1$),
 the improvement brought by the inclusion of dynamical localization corrections is more striking: not only 
the theory corrects the gross overestimate of the mobility (i.e. the value at $\omega \to 0$) 
implied by band theory, but it very accurately captures the whole frequency response, including the
emergence of a localization peak at $\omega\simeq 2J$ and the precise shape of the absorption 
tail at higher frequencies. The discrepancy observed at the low frequency absorption edge could 
instead be related to the small size of the cluster 
studied in Ref. \onlinecite{Fehske-PRB05}, which is limited to 6 sites.
As in the previous case, also here the choice $p=2.2\omega_0$ provides the overall best agreement with the exact result. 
We note that at this large level of disorder, the DLC result is essentially indistinguishable from the TL result (not shown).

\subsection{Breakdown of band transport}
\label{sec:breakdown}

\begin{figure}
   \centering
   \includegraphics[width=8.7cm]{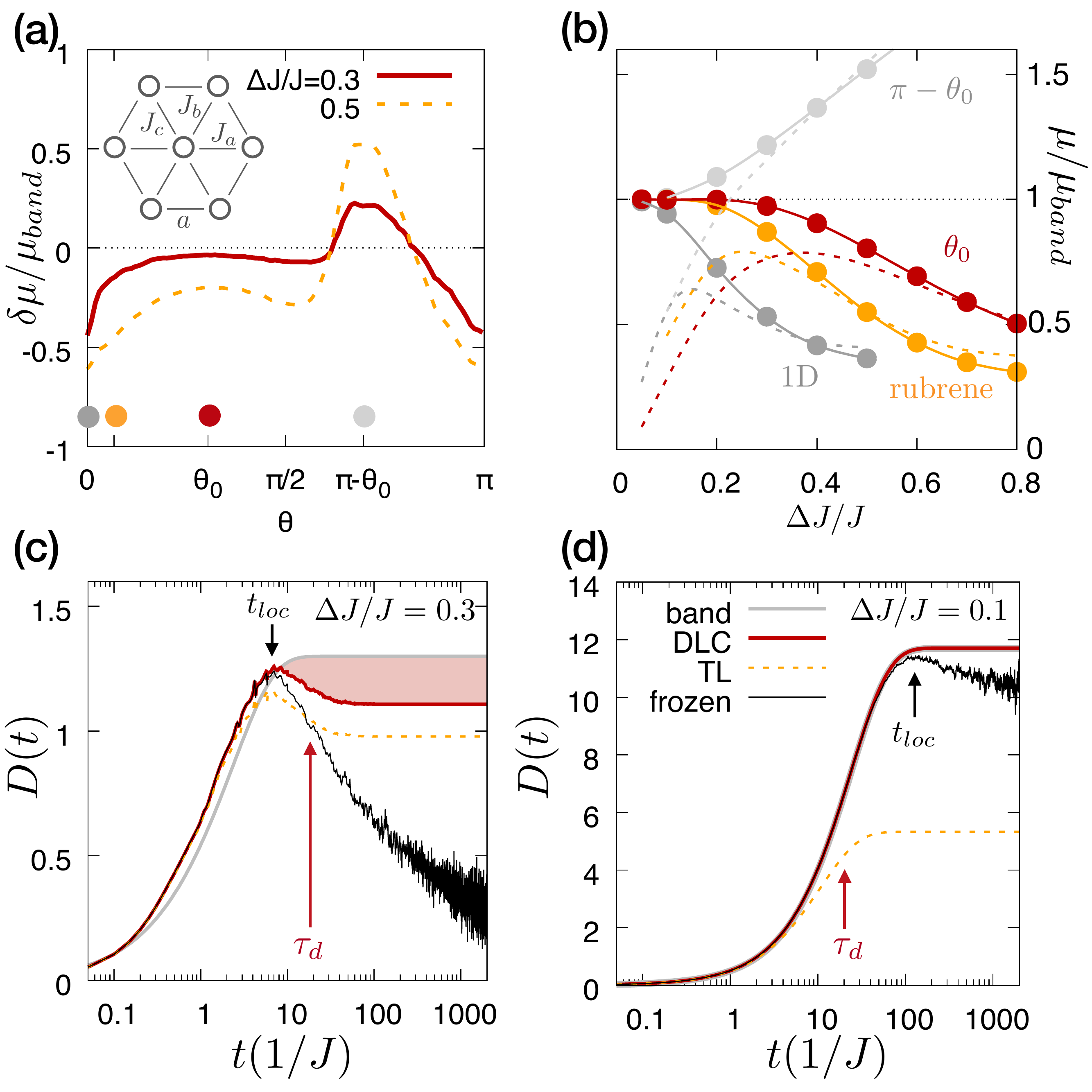}
 \caption{
 (a) Dynamical localization corrections $\delta \mu$
 for hole carriers on the tight-binding model defined in the text (sketched), 
 as a function of the band parameter $\theta$, calculated for  $\tau_{d}=20/J$ and $T/J=0.25$ (angular averaged).
%   Diffusion is in units of $a^2 J/\hbar$, the right axis is the corresponding mobility using $a=6\AA$.
% The sketch represents the hexagonal lattice studied in this work.
 (b) Mobility along the $a$ direction, scaled w.r.t. the band result ($\mu_{band}\propto (\Delta J/J)^{-2}$, 
 see Appendix \ref{sec:band})
 for the structures indicated by the labels ($\theta=0,0.21,\theta_0$ and $\pi-\theta_0$, see also symbols in panel a), 
 from TL (dashed lines) and DLC (full lines and symbols).
 (c) Time-dependent diffusivity calculated 
 for the band structure of rubrene\cite{NMat17}, $\theta=0.21$,  with  $\Delta J/J=0.3$.  Time is in units of $1/J$, 
 diffusivity is in units of $a^2 J/\hbar$.
 (d) same, with  $\Delta J/J=0.1$.}
\label{fig:diffcomp}
 \end{figure}

Having validated the theoretical approach, 
we now study the emergence of dynamical localization corrections in the broad class of organic semiconductors, 
by analyzing the ensemble of models described by Eq. (\ref{eq:H}).
Fig. \ref{fig:diffcomp}(a) illustrates the evolution of the quantum  
correction term $\delta \mu$ of Eq.  (\ref{eq:muRTE}) relative to the band value, 
as a function of the electronic structure parameter $\theta$.
Results are shown for two values of the energetic disorder, $\Delta J/J=0.3$ and $0.5$, and $p/J=0.05$ ($\tau_{d}=20/J$).
The quantum correction to the band mobility is 
predominantly negative
and expectedly increases in magnitude upon increasing the amount of disorder. 
The importance of processes beyond Bloch-Boltzmann theory is very much dependent 
on the band structure, which is in agreement with the general observation  
that different crystal structures are differently affected by disorder,
with isotropic two-dimensional bands 
being the most resilient to localization processes
\cite{NMat17}. 
The correction term is indeed maximum for 
one-dimensional structures ($\theta=0,\pi$), while it is minimized 
at the isotropic point ($\theta=\theta_0=\arccos(1/\sqrt{3})$),
where it becomes practically negligible at $\Delta J/J=0.3$.
The structures around $\pi-\theta_0$ are exceptions, exhibiting a  positive correction term: 
there, due to a negative combination of the signs of the transfer integrals, $J_aJ_bJ_c<0$,
a van Hove singularity of states resilient to localization 
arises close to the hole band edge. Thermal population of these states
causes the mobility to rise above the band value.

The detailed dependence of quantum corrections on the amount of disorder is illustrated in Fig. \ref{fig:diffcomp}(b), 
where we show the ratio of the total calculated mobility $\mu$ to the band prediction $\mu_{band}$, 
for selected electronic structures (solid lines and points). 
For each  structure, one can identify a  value of 
$\Delta J$ above which 
the mobility significantly deviates from the band value, signaling the emergence of quantum processes.
For example, band transport holds up to $\Delta J/J\simeq 0.3$ in the isotropic structure (red),
while it breaks down already at  $\Delta J/J\simeq 0.1$ in the one-dimensional case (dark gray). 
The neglect of quantum processes
beyond this point can lead to gross quantitative errors in the estimated mobility:
at the largest values of $\Delta J/J$ studied here, for example,  $\mu$ can deviate from the band value by 
 up to a factor of three.

In Fig. \ref{fig:diffcomp}(b) we also report the transient localization result, shown as dashed lines. 
The latter is applicable in the strong disorder regime, where it closely matches
the result of Eq. (\ref{eq:muRTE}).
Upon reducing the disorder strength, however, the TL result does not tend to the band limit as required, 
but instead it incorrectly bends downwards.
Band theory and TL theory therefore appear to be complementary, each addressing a different regime of parameters.  
Eq. (\ref{eq:muRTE}) allows to bridge continuously between these two limiting regimes.

\subsection{Localization corrections in the time domain}
\label{sec:timedomain}
The buildup of quantum localization processes can be  directly visualized 
by tracking the time-dependent diffusivity $D(t)$,
as shown in Fig. \ref{fig:diffcomp}(c) for $\Delta J/J=0.3$ and $\theta=0.21$ 
(these are the values of the microscopic parameters calculated for rubrene in Ref. \onlinecite{NMat17}).
The semiclassical diffusivity (gray) exhibits a monotonic evolution from ballistic at short times, $D\propto t$, to diffusive, $D\to \mathcal{D}_{SC}$ 
as $t\to \infty$, showing no hint of localization. Localization processes are instead fully developed 
when considering the exact evolution in a frozen molecular environment
(black thin line), as derived from the reference correlation function $C_0$ introduced in Sec. \ref{sec:dynloc}.
Their onset  can  be identified with the locus of the maximum of $D(t)$, that we denote as $t_{loc}$,
and  they are responsible for the subsequent steady decrease
of the diffusivity, which
vanishes at long times.
When the disorder is dynamic,
such localization corrections are initially retained, leading to a partial suppression of 
the diffusivity (red curve and hatched region) w.r.t. the band value. 
The suppression of $D(t)$ however stops at times 
$t\gtrsim \tau_{d}$,
which follows from the fact that 
its derivative $\delta C(t)$ vanishes
(cf. Eq. (\ref{eq:RTE})). 

By tracking the time evolution of the diffusivity, we can now better 
 visualize what controls the
emergence of dynamical quantum corrections in Eq. (\ref{eq:muRTE}). 
When the disorder is sufficiently strong (Fig.  \ref{fig:diffcomp}(c))
the condition $t_{loc}<\tau_{d}$ is fulfilled, and 
localization corrections can develop  \textit{before} they are suppressed by decorrelation due to the molecular dynamics. 
Reducing the disorder strength makes localization processes less efficient,
resulting in an increase of the localization time $t_{loc}$.
When the latter reaches $\tau_{d}$,
the quantum corrections cannot develop anymore, because they are cut off 
at their very onset by  decorrelation,  cf.  Eq. (\ref{eq:RTE}).  Quantum processes then become irrelevant
and band theory applies. Therefore,  the condition $t_{loc} \lesssim \tau_{d}$ 
is what marks the breakdown of semiclassical behavior.
According to this argument, the emergence of dynamical localization corrections is subject to the existence of either 
sufficiently   strong (low $t_{loc}$) or sufficiently slow (large $\tau_{d}$) disorder fluctuations.  
Both conditions are naturally realized in organic crystals, where large amplitude, slow intermolecular 
fluctuations arise owing to the large masses of the molecular constituents and to the weak intermolecular binding forces.

The disappearance of localization corrections at low disorder
is shown in Fig. \ref{fig:diffcomp}(d), where we report the diffusivity calculated for 
$\Delta J/J=0.1$, well within the band regime. 
In this case $t_{loc}>\tau_{d}$ and the full  diffusivity (red)
is essentially  indistinguishable from the band result, implying $\delta \mu\simeq 0$.

Finally, Fig.  \ref{fig:diffcomp}(d) also shows that applying TL theory when $t_{loc}>\tau_{d}$  
incorrectly underestimates the diffusion constant (dashed line),
which is at the origin of the non-monotonic behavior exhibited  in Fig. \ref{fig:diffcomp}(b).

\subsection{Localization corrections in the frequency  domain}

The conduction properties under a constant applied field
and the response of charge carriers  in the frequency domain are deeply intertwined.
In the band transport regime, the optical absorption 
exhibits a simple Lorenztian (Drude) shape, which is a monotonically decreasing function of frequency (cf. Appendix \ref{sec:band}).
Upon increasing the disorder strength, the suppression of the mobility (and hence the d.c. conductivity) by
localization processes
induces a dip in the absorption at $\omega=0$, as already shown in Fig. \ref{fig:validation}(b). 
This shifts the absorption maximum to finite frequencies
\cite{KavehMott,Smith,Ciuchi11,Fratini14,DeFilippis-PRL15}, providing 
a direct and measurable signature of the breakdown of band transport.

To assess how the emergence of dynamical localization corrections 
is reflected in the optical conductivity, 
we take advantage of the following  exact formula \cite{Ciuchi11} 
$ \sigma(\omega)=2 \tanh \frac{\beta \omega}{2} \int_0^\infty dt \sin (\omega t) D(t)$,
whose second derivative 
reads
\begin{equation}
\left.\frac{d^2\sigma}{d\omega^2}\right|_{\omega=0}=
- \beta \left[ 2\int_0^\infty \left[ \mathcal{D}-{D(t)}\right] t \ dt +\frac{\mathcal{D}\beta^2}{6} \right].
\label{eq:curvature}
\end{equation}
It is clear from the above relationship that whenever the diffusivity
is a monotonically increasing function of time, so that $D(t)$ is always lower than its long time limit $\mathcal{D}$,
the curvature is necessarily negative and the optical conductivity
remains peaked at $\omega=0$. This is the case in particular for the
band diffusivity shown in Figs. \ref{fig:diffcomp}(c) and (d),
in agreement with the resulting  Drude  behavior.
A necessary condition for the emergence of a finite-frequency peak is instead the existence of a region
where   $D(t)>\mathcal{D}$, as indeed happens when quantum corrections are relevant 
(red curve in Fig. \ref{fig:diffcomp}(c)).
This, together with the analysis of the time-dependent diffusivity given in the preceding paragraphs,
shows that the emergence of a dip in the optical
absorption essentially coincides with the crossover condition $\tau_{d}\sim t_{loc}$. The  
change of sign of Eq. \ref{eq:curvature} can therefore be used to identify  
the breakdown of the band picture. 
The equivalence between these two conditions is further explored next.

\begin{figure}[h]
   \centering
   \includegraphics[width=8.8cm]{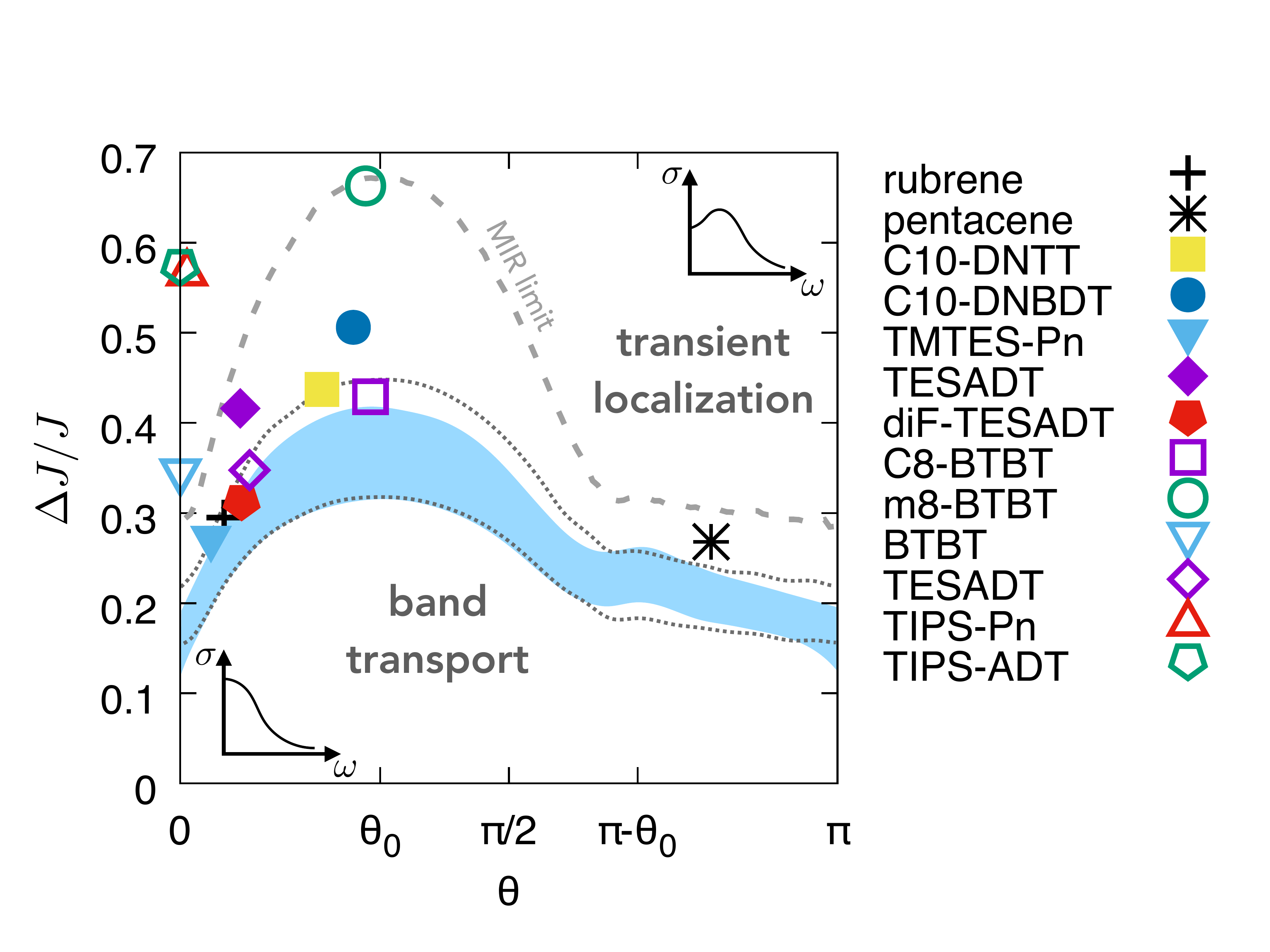}
  \caption{Regimes of room temperature charge transport in two-dimensional organic semiconductors. 
  The shaded area denotes the emergence of dynamical localization corrections as determined 
  from Eq. (\ref{eq:curvature}). The upper and lower edge are determined by choosing   $\tau_d=10/J$ and  
  $20/J$ respectively, both with $J=0.1$eV  and $T=0.25 J$. The dotted lines are the corresponding estimates from Eq. (\ref{eq:qualcrit}), and the
  dashed line is the MIR limit (see text).
    Data points are from Refs. \onlinecite{NMat17} (full symbols) and \onlinecite{Harrelson} (open symbols). 
    The insets illustrate the
  characteristic shapes of the optical absorption in the two different regimes.}
 \label{fig:PD}
 \end{figure}

\subsection{Transport phase diagram}
\label{sec:PD}
The locus of the breakdown of band transport as determined from the condition $d^2\sigma/d\omega^2|_{\omega=0}=0$
is reported in Fig. \ref{fig:PD} as a function of the electronic structure parameter $\theta$ (shaded area).
The symbols locate the electronic structures  and disorder levels calculated for a
number of high-mobility organic semiconductors of current interest  \cite{NMat17,Harrelson}.
For completeness we have included materials that either exactly
fulfill the condition $J_b=J_c$, or are sufficiently close to it.
The shaded area delimits the crossover lines calculated for values of the molecular fluctuation time 
$\tau_{d}$ comprised in the interval $10-20 \ \hbar/J$, which is the typical range  encountered in 
materials.
Remarkably,  all the reported compounds are characterized by 
sizable dynamical localization processes, making band transport theory inappropriate.  Several of them, however,
are located very close to the crossover region.
In the case of rubrene, in particular,
this is in agreement with the fact that a finite-frequency peak is 
observed in optical absorption experiments at room temperature\cite{Fischer,Li},
yet  a normal Drude shape is recovered upon reducing the amount of thermal disorder \cite{Okamoto}. We stress that the reported crossover line corresponds to the
ideal situation where the only source of randomness is from transfer integral fluctuations. In real conditions, local site-energy fluctuations
originating from  the coupling to intra-molecular vibrations as well as extrinsic sources of disorder will likely shift the crossover to lower values of $\Delta J/J$.

At any rate, we observe that while low levels of disorder and isotropic band structures have been
independently achieved in current materials, no compound exists yet that is able
to combine such optimal features together. If such a compound could be
synthesized, we argue that it would fully enter the
%overcome the crossover region, entering the
band transport regime, possibly opening new perspectives for organic-based applications.

\section{Concluding remarks}
\label{sec:MIR}

Poor  conduction properties are  commonly observed 
in  broad and diverse classes of solids, 
including disordered and amorphous metals, polymers,
materials with strong electron-phonon interactions and strongly correlated electron systems. 
Regardless of the microscopic origin, all these "bad" conductors and semiconductors
have in common a fundamental breakdown of the weak-scattering hypothesis underlying band transport theory.
In this respect, we speculate that the mechanisms described in this work 
could generally apply 
to other classes of materials, where slowly fluctuating degrees of freedom of electronic, magnetic or
vibrational origin are coupled to the carrier motion.
While the microscopic analysis of specific cases is clearly beyond the scope of this work, 
we can take advantage of the physical insight gained here 
to derive a simple phenomenological formula that will be of practical use to 
assess the presence of dynamical localization corrections in general situations.

To this aim we go back to Sec. \ref{sec:timedomain}, where we have shown that
localization processes  arise
when the localization time is shorter
than the velocity decorrelation time, i.e 
\begin{equation}
t_{loc} \lesssim \tau_d.
\label{eq:DMCcrittimes}
\end{equation}
We observe that this typically occurs in a regime where the level of disorder
is sizable on the scale of the band energy (cf. Fig. \ref{fig:diffcomp}). 
In this regime, the localization time $t_{loc}$ scales with the semiclassical scattering rate $\tau$ within a numerical factor,
and the two merge in the strong disorder limit, where $t_{loc}\simeq \tau$. 
Specifically, we have checked from the data of Fig. \ref{fig:diffcomp} that even in 
the isotropic 2D case, where localization corrections are the weakest, 
the ratio $ t_{loc}/ \tau$ falls below 2 as soon as $\Delta J/J \gtrsim 0.4$. In 1D, this is attained 
already at $\Delta J/J \gtrsim 0.2$.

It is then appealing to rewrite Eq. (\ref{eq:DMCcrittimes}) by replacing the (quantum) 
localization time $\tau_{loc}$ with the more familiar
(semiclassical) scattering time $\tau$. We therefore write
\begin{equation}
\frac{1}{\tau}\simeq \eta\frac{1}{\tau_d} \; \; \; \mathrm{DMC}
\label{eq:qualcrit}
\end{equation}
with $\eta$ a numerical prefactor. 
As a consistency check for Eq. (\ref{eq:qualcrit}), when $\tau_d\to \infty$ 
we recover the known result that in the low density limit appropriate to non-degenerate semiconductors, 
localization corrections are always relevant when disorder is static \cite{RMP84}.
To fix the value of $\eta$, in Fig. \ref{fig:PD} we compare Eq. (\ref{eq:qualcrit}) with the rigorous condition
obtained from Eq. (\ref{eq:curvature}), for two different values of the fluctuation time, 
$\tau_d=10/J$ and  $20/J$. We find that a very good match 
is obtained for $\eta\simeq 4$ (shaded area and dotted lines respectively). 
The criterion Eq. (\ref{eq:qualcrit})  can now be straightforwardly  applied to a variety physical situations of interest,
provided that  $\tau_d$ is identified with the timescale of the relevant degrees of freedom that couple to the electron motion.

It is interesting to compare this result with the phenomenological Mott-Ioffe-Regel (MIR)  criterion,
which signals the disappearance of Bloch states. 
In its original formulation \cite{Ioffe-Regel}, this is identified 
as the point where the semiclassical mean-free-path becomes comparable to the lattice spacing, i.e. $\ell \simeq a$.
Using $\ell^2=\langle v^2 \rangle \tau^2$,
this can be rewritten in terms of the scattering rate as 
\begin{equation}
\frac{1}{\tau} \simeq \frac{\langle v^2 \rangle^{1/2}}{a} \; \; \; \mathrm{MIR}.
\label{eq:MIRorig}
\end{equation}
This condition is illustrated in Fig. \ref{fig:PD} as a dashed line, and it is located well above the crossover region.
The fact that deviations from band theory arise
before the MIR limit is reached
is consistent with the fact that the buildup of quantum localization corrections requires 
the existence of coherently propagating Bloch waves to start with.
The comparison of Eqs. (\ref{eq:qualcrit}) and (\ref{eq:MIRorig}) shows that this can only happen
in systems where the disorder dynamics is sufficiently slow,  as is the case here.

Finally, we note that a slightly different formulation of the MIR criterion
is often used\cite{Calandra,Hussey}, which predicts the disappearance of Bloch states 
when the semiclassical scattering rate $1/\tau$ reaches a 
fraction of the bandwidth, i.e.
\begin{equation}
\frac{1}{\tau}\simeq \xi \frac{J}{\hbar} \; \; \; \mathrm{MIR \ (spectral)}
\label{eq:MIRcrit}
\end{equation}
with $\xi$  again a numerical factor. 
The two conditions Eq. (\ref{eq:MIRorig}) and Eq. (\ref{eq:MIRcrit})  become equivalent in the ultra-high temperature limit 
$T\gtrsim J$, in which case $\langle v^2 \rangle^{1/2}/a\propto J/\hbar$ independent on temperature (the average velocity 
entering in Eq. (\ref{eq:MIRcrit}) instead recovers instead
$\langle v^2 \rangle\propto k_BT/m^*$ at low temperature).
Comparing  Eq. (\ref{eq:MIRcrit}) and Eq. (\ref{eq:qualcrit}), we again conclude 
that for sufficiently slow vibrations, i.e. $J\gg \omega_0$, 
the dynamical localization corrections emerge in a region where Bloch states are not yet suppressed.

\appendix

\section{Band transport theory}
\label{sec:band}
Here we start from the textbook equations of Bloch-Boltzmann band transport theory to evaluate the time-dependent  
diffusivity $D_{SC}(t)$ and anticommutator correlation function $C_{SC}(t)$  needed in the main text, and provide
useful analytical formulas for the mobility in different models of electron-phonon coupling that are relevant to 
high-mobility organic semiconductors.

\subsection{Time-dependent diffusivity}

In band transport theory, one starts with Bloch states in the periodic molecular lattice, having
momentum $k$, energy $\epsilon_k$ and velocity $v_k$ in a given direction. The diffusion constant of each 
band state is $\mathcal{D}_k=v_k^2 \tau_k$, with $\tau_k$ a decay time determined by scattering off disorder and
lattice vibrations (see Section \ref{sec:models} below).  The mobility is then obtained as 
$\mu_{band}=e \beta  \mathcal{D}_{SC}  = \beta e \langle v_k^2\tau_k \rangle$, where  
the symbol $\langle \cdots \rangle$ indicates thermal averaging over the states and $\beta=1/k_BT$, with $k_B$ the 
Boltzmann constant.

The time-dependent diffusivity is related to the real part of the frequency-dependent 
conductivity $\sigma(\omega)$ per particle via the following expression \cite{Ciuchi11}
\begin{equation}
D(t)=\frac{1}{\pi}  \int_0^\infty \frac{\sin(\omega t)}{\tanh(\beta \omega/2)} \sigma(\omega) d\omega,
\label{eq:diffsigma}
\end{equation}
which is exact for non-degenerate carriers.
The optical conductivity per band carrier is given by \cite{Allen}
\begin{equation}
\sigma_{SC}(\omega)= \frac{\beta}{Z}  \mathrm{Re} \sum_k \frac{ v_k^2\tau_k}{1-i \omega \tau_k} e^{-\beta \epsilon_k},
\label{eq:optcondB}
\end{equation}
with $Z=\sum_k e^{-\beta \epsilon_k}$ the partition function.
The corresponding diffusivity 
is obtained via direct integration of Eq. (\ref{eq:diffsigma}). 
As Eq. (\ref{eq:optcondB}) is a 
linear superposition, the contributions from individual states can be separated as $D_{SC}(t)=(1/Z)\sum_k e^{-\beta \epsilon_k} D_k(t)$, with 
\begin{equation}
D_k(t)= \frac{\beta}{\pi} \mathrm{Re} \int_0^\infty \frac{\sin(\omega t)}{\tanh(\beta \omega/2)} \frac{ v_k^2\tau_k}{1-i \omega \tau_k} d\omega.
\label{eq:Dksingle}
\end{equation}
The integral can be performed via contour integration, leading to
\begin{equation}
D_k(t)= {v_k^2 \tau_k}\left \lbrack 
1 -\frac{\beta}{2\tau_k}\frac{e^{-t/\tau_k}}{\tan (\beta/2\tau_k)}
+\sum_{n>0} \frac{2e^{-\omega_n t}}{1-(\omega_n\tau_k)^2} \right\rbrack
\label{eq:fullDB}
\end{equation}
where we have introduced the bosonic Matsubara frequencies
$\omega_n=2\pi n/\beta$. The result of the last summation has a closed expression 
in terms of the Hurwitz-Lerch transcendent function $\Phi$. Denoting $z=e^{-2\pi t/\beta}$, we have 
$\sum_{n>0} \frac{2e^{-\omega_n t}}{1-(\omega_n\tau_k)^2}=[z\beta/\pi (\tau_k)^2]
(\Phi(z,1,1+\omega_1 \tau_k)-\Phi(z,1,1-\omega_1 \tau_k))$.
The anticommutator velocity-velocity correlation function $C_{SC}(t)$ can be straightforwardly obtained from 
Eq. (\ref{eq:fullDB}) using the definition $C_{SC}(t)=2d D_{SC}/dt$.

In the high temperature/weak scattering limit, $T\gg 1/\tau_k$, the explicit sum over Matsubara frequencies 
drops out and the above expression simplifies to 
$D_k(t)\simeq v_k^2\tau_k [1-e^{-t/\tau_k}]$ (this result can be obtained straightforwardly from 
Eq. (\ref{eq:Dksingle}) by taking the classical limit for the detailed balance factor, $\tanh(\beta \omega/2)\to \beta \omega/2$). 
This form corresponds to a simple exponential decay of the velocity correlation 
function \cite{Ciuchi11}. It describes ballistic behavior at short times, 
$D_k\sim v_k^2 t$, followed by diffusive behavior  $D_k\to const = v_k^2 \tau_k$ at times $t\gg \tau_k$. 
The diffusivity in this case is a monotonically increasing function of time. 
According to the full expression Eq. (\ref{eq:fullDB}), however, an anomalous time dependence
 can arise when the scattering rate becomes much larger than the thermal energy, in which case
deviations from the simple monotonic form above can appear. 
This happens because the ballistic velocity at short times becomes larger than $v_k$. 
Due to this initial overshoot, $D_k(t)$ reaches values larger than the long-time diffusion constant, 
which is then attained after going through a maximum. 
Such non-monotonic behavior arises when 
$1/\tau_k\gtrsim 3.15 T$. 
Note that, because the scattering rate for classical vibrations 
increases with $\sqrt {T}$ due to the equipartition principle,  a regime
of anomalous diffusivity can in principle be attained at low temperature.

\subsection{Average scattering time}
\label{sec:avgtau}
The full electrodynamic response of non-degenerate carriers in the Bloch-Boltzmann approximation,
as given by Eq. (\ref{eq:optcondB}), is a superposition of Lorentzians of widths $1/\tau_k$, weighted
by the corresponding thermal factors. This, in principle,
deviates from a simple Lorentzian shape, as would be predicted instead within Drude theory.
%, and it is not possible to provide a simple expression in general. 
A simpler approximation for $\sigma(\omega)$ \cite{Allen}, and therefore for $D_{SC}(t)$, is obtained by introducing a
\textit{single}, k-independent relaxation time $\tau$. The latter is univocally determined from the knowledge of 
the long-time diffusivity, as
\begin{equation}
\tau=\langle v_k^2 \tau_k \rangle/\langle v_k^2\rangle
\label{eq:tauavg}
\end{equation} 
which is the proper thermal average of the transport scattering 
time $\tau_k$, as defined in Ref. \cite{Allen}. By construction, the above equation recovers 
the correct diffusivity 
$\mathcal{D}_{SC}=\langle v_k^2\rangle \tau=\langle v_k^2 \tau_k \rangle$ and mobility $\mu_{band}=e\beta \langle v_k^2\rangle \tau$.
The corresponding optical conductivity then takes the simple Drude form
\begin{equation}
\sigma(\omega)=\frac{\sigma_0}{1-i\omega \tau}
\label{eq:optavg}
\end{equation} 
with $\sigma_0=\beta \langle v_k^2 \tau_k \rangle$.
In all cases studied the time-dependent diffusivity and optical conductivity obtained from the average $\tau$ are 
either very close to or  indistinguishable from those obtained from the full k-dependent expressions.
We therefore use the former simplified framework for the evaluation of the quantum corrections in the main text.

\subsection{Calculations on specific models}
\label{sec:models}
Let us consider the scattering of a k-state off phonon modes 
of momentum $q$ and frequency $\omega_0$, as described by the interaction Hamiltonian
\begin{equation}
H_I=(1/N)\sum_{k,q} \sum_{\delta} \alpha_{k,q}^{(\delta)} c^+_{k+q}c_k x_{q,\delta}.
\label{eq:HI}
\end{equation} 
Here 
$N$ is the number of molecules, $c^+_k,c_k$ the creation and annihilation operators for carriers, 
$x_{q,\delta}$ the deformation mode corresponding to a given bond direction $\delta$, and
we  set $\hbar=1$. Straightforward algebra allows to write 
the interaction matrix elements for  uncorrelated bond disorder as
$[\alpha_{k,q}^{(\delta)}]^2=4 \alpha_\delta^2 [\cos ((k+q/2)\cdot \delta)]^2$, 
with $\delta$ the vectors connecting nearest-neighbours as shown in Fig. 1a and 
$\alpha_\delta=dJ_\delta/dx_\delta$ the sensitivity
of the transfer integrals to intermolecular deformations. 
Upon substituting this expression, Eq. (\ref{eq:HI}) becomes equivalent to Eq. (1) of the main text,
now expressed in momentum space.
The canonical 2nd quantization expression for the electron-phonon interaction is obtained by expressing
the bond coordinate in terms of dimensionless bosonic operators as 
$x_{q,\delta}=(\omega_0/2K)^{1/2}(b_{q,\delta}^+ + b_{q,\delta})$ with $K$ the spring constant, 
so that the
electron-phonon coupling matrix element becomes $g_{k,q}^{(\delta)}=(\omega_0/2K)^{1/2}\alpha_{k,q}^{(\delta)}$,
and correspondingly $g_\delta=(\omega_0/2K)^{1/2}\alpha_\delta$.
As is customarily done, we  introduce  a set of dimensionless coupling parameters 
$\lambda_\delta=\alpha_\delta^2/(4J_\delta K)=(g_\delta^2/\omega_0)/2J_\delta $.
The classical (thermal) 
fluctuation of the transfer integrals can be written as 
$(\Delta J_\delta)^2 = \alpha_\delta^2 T/K =2 \lambda_\delta J_\delta T$ using the 
equipartition principle. 
In general, one can define global parameters $J^2=\sum_\delta J_\delta^2$  
and $\Delta J^2=\sum_\delta \Delta J_\delta^2$.
In Sec. \ref{sec:breakdown} we 
consider a model where the relative fluctuations in the different bond directions are all equal, 
i.e. $\Delta J_\delta/J_\delta=\Delta J/J$ for all $\delta$. This 
corresponds to the choice of an isotropic coupling $\lambda_\delta\equiv \lambda$, 
and leads to $(\Delta J)^2 =2\lambda J T$ in the thermal fluctuation regime.

Other models of interest can be put in the form of Eq. (\ref{eq:HI}). 
In the case of fully correlated bond disorder, as studied in Refs. \cite{Fratini09,Fratini-AFM16}, 
the matrix element 
reads $[\alpha_{k,q}^{(\delta)}]^2=4 \alpha_{SSH}^2 [\sin ((k+q)\cdot \delta)-\sin (k\cdot \delta)]^2$, with now
$(\Delta J_{SSH})^2 = 2\alpha_{SSH}^2 T/K=4 \lambda_{SSH} J T$.
The prefactor $4$ instead of $2$ arises from the fact that the fluctuation of the transfer integral now
arises from two independent modes located on adjacent sites. 
Finally, diagonal (intra-molecular) electron-phonon interactions  correspond 
to a constant $\alpha_{k,q}=\alpha_H$, where $\alpha_H$ measures the variation of 
the local molecular energy level with respect to an intra-molecular deformation $x$, and correspondingly
$\Delta ^2 = \alpha_H^2 T/K=2 \lambda_H J T$.

The momentum scattering rate is evaluated in $d$ dimensions as 
\begin{eqnarray} \nonumber
& & \frac{1}{\tau_k}= 2\pi \int \frac{d^dq}{(2\pi)^d} \; g^2_{k,k+q}  F_{k,k+q}
 \times \\
  \nonumber &  & \times
  \left\lbrace
    \left[n_b+f_{k+q}\right]
    \delta(\epsilon_{k}-\epsilon_{k+q}+\omega_0) \right.\\
& &  + \left. \left[n_b+1-f_{k+q}\right]
    \delta(\epsilon_{k}-\epsilon_{k+q}-\omega_0)
\right\rbrace   \label{eq:rate}
\end{eqnarray}
with $n_b$ the phonon population and $f_{k+q}$ the Fermi occupation of the final state $k+q$,
which can be set to zero at low carrier densities, and we have introduced the compact notation 
$g^2_{k,k+q}=\sum_\delta [g^{(\delta)}_{k,k+q}]^2$.
The two terms between brackets  originate from phonon emission and absorption 
respectively. The geometric factor $F_{k,k+q}= 1- v_k\cdot v_{k+q}/v_k^2$ 
measures the loss of momentum occurring at each scattering event, thereby differentiating the transport scattering time
from the quasiparticle scattering time, which is instead obtained by setting $F_{k,k+q}= 1$. 
The form of $F_{k,k+q}$ used here is the proper generalization to generic
band structures of the textbook expression $-k\cdot q/k^2$, which
only applies to isotropic, parabolic band dispersions. Note that the factor $F_{k,k+q}$ is often omitted in practical calculations
\cite{NMat17,Bernardi}, generally leading to quantitatively incorrect values for the mobility (see below).

In the quasi-elastic limit where the intermolecular
vibration frequencies set the smallest energy scale in the problem, 
$\omega_0\ll T,J$ the  scattering time simplifies to
\begin{equation}
  1/\tau_{k}= \frac{2k_BT}{\hbar \omega_0}  2\pi\int \frac{d^dq}{(2\pi)^d} \; g^2_{k,k+q} F_{k,k+q}
 \; \delta(\epsilon_k-\epsilon_{k+q}).
 \label{eq:tauclass}
\end{equation}
From this result and from the expressions of $g^2_{k,k+q}$ given above it is clear that 
in all models considered here, the band mobility at any given temperature in the classical regime $T\gtrsim \omega_0$ 
scales with disorder strength as $\mu\propto \Delta J^{-2}$, 
which reflects the second-order nature of the scattering process.

\subsection{Analytical results in 1D}

For charge carriers in one space dimension, most calculations can be 
performed analytically. In what follows we express distances 
in units of the lattice spacing $a$, we set $\hbar=1$ and mobility units to $\mu_0=ea^2/\hbar$.
Taking $J_a=J$ and $J_b=J_c=0$, we have 
$\epsilon_k=-2J\cos(k)$ and $v_k=2J\sin(k)=\sqrt{4J^2-\epsilon_k^2}$.  
The density of states is $\rho(\nu)=1/\sqrt{4J^2-\nu^2}/\pi$ and we have
\begin{eqnarray}
Z&=&I_0(2\beta J) \\
\langle v_k^2 \rangle &=& 2J^2 \phantom{x}_0F_1(2,(\beta J)^2)/Z.
\end{eqnarray}
The last equation allows to determine the average scattering time from the knowledge of the diffusion constant $\mathcal{D}$,
via Eq. (\ref{eq:tauavg}).
The scattering time, diffusion constant and charge mobility 
can be calculated from Eq. (\ref{eq:tauclass}) in the following cases.

\paragraph{Uncorrelated bond fluctuations $\Delta J$,} i.e. off-diagonal thermal disorder: 
\begin{eqnarray}
\tau_k&=&\left(\frac{J}{\Delta J}\right)^2      \frac{(4J^2-\epsilon_k^2)^{1/2}}{8 J^2} \label{eq:tauuncorrvtx}\\
\tau_k^{qp}&=&\left(\frac{J}{\Delta J}\right)^2 \frac{(4J^2-\epsilon_k^2)^{1/2}}{4 J^2+\epsilon_k^2} \label{eq:tauuncorr}
\end{eqnarray}
where the last expression is the quasiparticle scattering time, obtained by setting $F_{k,k+q}=1$ in Eq. (\ref{eq:tauclass}). 
By performing the momentum integrations
we obtain
\begin{eqnarray}
\mathcal{D}&=&\frac{2\beta J\cosh(2\beta J)-\sinh(2\beta J)}{2\pi \beta^3 (\Delta J)^2 Z} \label{eq:DB1D}\\
\mu&=&\mu_0  \frac{2\beta J\cosh(2\beta J)-\sinh(2\beta J)}{2\pi (\beta J)^2 (\Delta J/J)^2 I_0(2\beta J)} \\
&\simeq &   \mu_0   \frac{(T/J)^{-1/2}}{2\pi^{1/2} \lambda} \label{eq:muB1DloT}\; \; \; T \ll J
\label{eq:muB1D}
\end{eqnarray}
where $I_0$ and $\phantom{x}_0F_1$ denote respectively the modified Bessel function of the first kind and the 
regularized hypergeometric function. In the last equation, we have introduced  
$(\Delta J/J)^2=2 \lambda T/J$ using the definition 
of the dimensionless electron-phonon coupling $\lambda=\alpha_\delta^2/(2KJ)$. 

\paragraph{Diagonal disorder $\Delta$.} In this case we have
\begin{eqnarray}
\tau_k&=&\left(\frac{J}{\Delta}\right)^2 \frac{(4J^2-\epsilon_k^2)^{1/2}}{2J^2}\\
\tau_k^{qp}&=&\tau_k \label{eq:taudiag}\\
\mathcal{D}&=&\frac{2\beta J\cosh(2\beta J)-\sinh(2\beta J)}{(\pi/2) \beta^3 \Delta^2 Z}\\
\mu&=&\mu_0 \frac{2\beta J\cosh(2\beta J)-\sinh(2\beta J)}{(\pi/2) (\beta J)^2 (\Delta/J)^2 I_0(2\beta J)}\\
&\simeq &   \mu_0   \frac{2(T/J)^{-1/2}}{\pi^{1/2} \lambda_H}\; \; \; T \ll J
\end{eqnarray} 
where we have used  $(\Delta /J)^2=2 \lambda_H T/J$.
Note that the transport scattering rate is formally equivalent to that arising from uncorrelated bond disorder 
(this equivalence is instead lost for the quasiparticle scattering time).  
As a result,  the mobility has exactly the same functional form as in Eqs. (\ref{eq:DB1D}-\ref{eq:muB1D}), but with a 
reduced prefactor: for a given energetic spread $\Delta$, diagonal disorder appears to be 4 times less 
effective than off-diagonal disorder $\Delta J$ in limiting the charge mobility, which is therefore 4 times larger. 
This result, which is exact in one dimension, remains approximately true in all
the two-dimensional structures studied here.

\paragraph{Correlated bond fluctuations $\Delta J$.} In this case we have
\begin{eqnarray}
\tau_k&=&\left(\frac{J}{\Delta J}\right)^2   \frac{1}{8(4J^2-\epsilon_k^2)^{1/2}} \label{eq:taucorrvtx} \\
\tau_k^{qp}&=& 2 \tau_k \label{eq:taucorr} \\
\mathcal{D}&=&\frac{\sinh(2\beta J)}{4\pi (\Delta J)^2 \beta Z}.\\
\mu&=& \mu_0    \frac{\sinh(2\beta J)}{4\pi (\Delta J/J)^2  I_0(2\beta J)} \label{eq:muBcorr1D}\\
&\simeq &   \mu_0   \frac{(T/J)^{-3/2}}{16\pi^{1/2}  \lambda}\; \; \; T \ll J \label{eq:muBcorr1DloT}
\end{eqnarray}
with $\lambda=\alpha_\delta^2/(2KJ)$, so that $(\Delta J/J)^2=4 \lambda T/J$ (see above).

\paragraph{Summary.}
\begin{figure}[h]
   \centering
      \includegraphics[width=8cm]{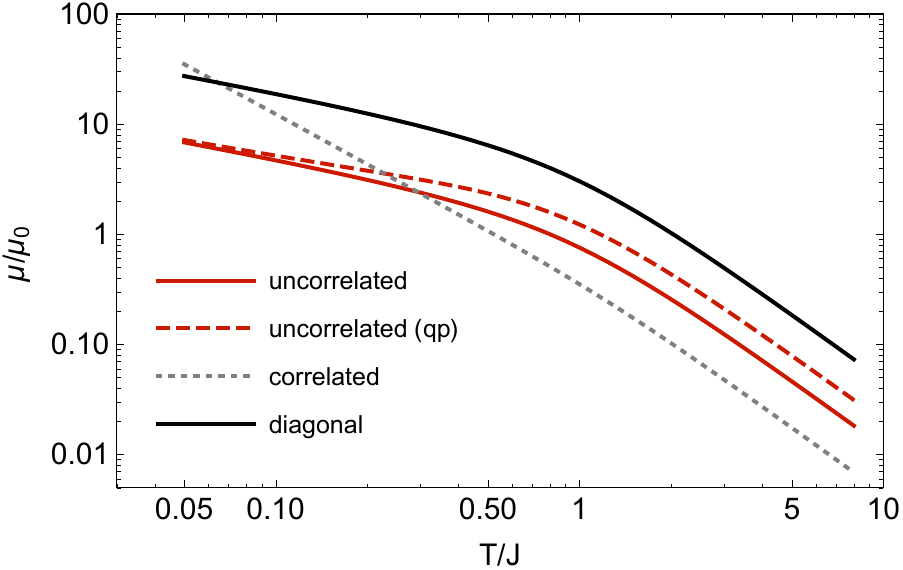}
  \caption{Band mobility vs temperature for charge carriers in 1D 
  in the quasi-static limit for phonons, according to different electron-phonon coupling 
  models. In all cases the dimensionless coupling strength $\lambda$ is chosen such that 
  $\Delta J/J=0.3$ at room temperature ($T/J=0.25$ having assumed $J=0.1$eV). 
  Mobility units are set by $\mu_0=ea^2/\hbar$, typically in the range $5-8$ cm$^2$/Vs.
  }
 \label{fig:muB}
 \end{figure}

Fig. \ref{fig:muB} illustrates the temperature dependence of the mobility obtained, 
in the quasi-static limit, in the different cases studied 
above. We first observe that the effect of the geometrical factor $F_{k,k+q}$ is rather small for 
uncorrelated bond disorder, especially at low temperature, and it is strictly irrelevant in the case of diagonal disorder, 
cf. Eq. (\ref{eq:taudiag}). 
This is in contrast with the case of correlated bond disorder, where its neglect 
leads to an overestimate of the mobility by a factor of 2 [cf. Eq. (\ref{eq:taucorr})], 
as already reported in Ref. \cite{Fratini09}. 
Second,  the temperature dependence for uncorrelated bond disorder in one dimension (and, similarly, for diagonal disorder)
is rather weak in the relevant temperature range around and below room temperature. At $T/J= 0.25$ the 
temperature exponent of the mobility $\mu\propto T^{-\gamma}$  is $\gamma=0.66$, and it tends to 
$\gamma=0.5$ when $T\ll J$, cf. Eq. (\ref{eq:muB1DloT}).  The exponent is slightly reduced,
$\gamma=0.46$, if the geometrical vertex factor is neglected, as reported in Fig. 3a of Ref.\cite{NMat17}, again recovering
$\gamma=0.5$ when $T\ll J$.  The temperature exponent is larger
for correlated bond disorder,  $\gamma\simeq 1.5$,  cf. Eq. (\ref{eq:muBcorr1DloT}), which holds for all $T\lesssim J$. 
We note that the mobilities calculated for correlated and uncorrelated bond disorder have 
accidentally very similar values at room temperature.

\subsection{Quantum phonons}
In all  cases reported above we have considered the quasi-elastic limit $\hbar \omega_0 \ll k_BT$, 
by setting explicitly $\omega_0\to 0$ in Eq. (\ref{eq:tauclass}). We illustrate 
here how the quantum nature of the phonons
is restored in the specific case of correlated disorder, by using the full expression Eq. (\ref{eq:rate}). 
The calculations 
%proceeds in a similar way for 
can be straightforwardly extended to the other cases considered above.

For each of the two processes described in  Eq. (\ref{eq:rate}), i.e. phonon emission and absorption, 
the delta function yields two solutions 
\begin{equation}
  \label{eq:deltas}
  \delta(\omega\pm \omega_0-\epsilon_{k+q})
  =\frac{\delta(q-q_+)+\delta(q-q_-)}{\sqrt{4J^2-(\epsilon_k\pm \omega_0)^2}} 
\end{equation}
with  
\begin{equation}
  \label{eq:qpm}
  k+q_{\pm}= \pm \kappa \;\; \;\;  ;  \;\;  \;\;\kappa=
\arccos \left[-\frac{\epsilon_k\pm \omega_0}{2J}\right].
\end{equation}
The corresponding  squared matrix element $g^2_{k,k+q}$ is equal to  $4g^2$ times 
\begin{equation}
  \label{eq:matrel+}
  1+\sin^2k -\left(\frac{\epsilon_k\pm \omega_0}{2t}\right)^2 - 2\sin k \sqrt{1-\left(\frac{\epsilon_k\pm \omega_0}{2t}\right)^2} 
\end{equation}
at $q=q_+$, and
\begin{equation}
  \label{eq:matrel-}
  1+\sin^2k -\left(\frac{\epsilon_k\pm \omega_0}{2t}\right)^2 + 2\sin k \sqrt{1-\left(\frac{\epsilon_k\pm \omega_0}{2t}\right)^2} 
\end{equation}
at $q=q_-$.
%, where we have defined $g^2=2\lambda J\omega_0$. 
In the absence of vertex corrections, i.e. setting $F_{k,k+q}= 1$, 
%all the other factors do not depend on $q$, then 
the square root term cancels from the sum, leading to:
\begin{eqnarray} \nonumber
& &   \frac{1}{\tau_k^{qp}} = 
    8g_{SSH}^2 \left\lbrace \frac{n_b+f(\epsilon_k+\omega_0)}{\sqrt{4J^2-(\epsilon_k+ \omega_0)^2}}
     \left[  2-\frac{\epsilon_k^2}{4J^2} -\frac{(\epsilon_k+
           \omega_0)^2}{4J^2}\right] \right.\\
&& +   \left.\frac{n_b+1-f(\epsilon_k-\omega_0)}{\sqrt{4t^2-(\epsilon_k- \omega_0)^2}}
     \left[   2-\frac{\epsilon_k^2}{4J^2} -\frac{(\epsilon_k-
           \omega_0)^2}{4J^2}\right] \right\rbrace.
   \label{eq:rateSSHqp}
\end{eqnarray}

Including the geometrical vertex $F_{k,k+q}= 1-v_k\cdot v_{k+q}/v_k^2$ restores the
square root terms in Eqs. (\ref{eq:matrel+}) and (\ref{eq:matrel-}). After some elementary algebra 
the transport scattering time is then obtained as
\begin{eqnarray} \nonumber
& &   \frac{1}{\tau_k} = 
    8g_{SSH}^2 \left\lbrace\frac{n_b+f(\epsilon_k+\omega_0)}{\sqrt{4J^2-(\epsilon_k+ \omega_0)^2}}
     \left[  4-\frac{\epsilon_k^2}{4J^2} -3\frac{(\epsilon_k+
           \omega_0)^2}{4J^2}\right] \right.\\
&& +  \left. \frac{n_b+1-f(\epsilon_k-\omega_0)}{\sqrt{4t^2-(\epsilon_k- \omega_0)^2}}
     \left[   4-\frac{\epsilon_k^2}{4J^2} -3\frac{(\epsilon_k-
           \omega_0)^2}{4J^2}\right] \right\rbrace.
   \label{eq:rateSSH}
\end{eqnarray}
The calculation of the diffusion constant $\cal{D}$ and of the mobility $\mu$ now proceed as in the
classical case. The energy integrals $\langle v_k^2 \tau_k\rangle$ cannot be cast in closed analytical form
and must be performed numerically.

We report for completeness the expressions obtained for uncorrelated disorder, i.e.
\begin{eqnarray} \nonumber
&&  \frac{1}{\tau_k^{qp}} = 
    4g_\delta^2\left\lbrace \frac{n_b+f(\epsilon_k+\omega_0)}{\sqrt{4J^2-(\epsilon_k+ \omega_0)^2}}
     \left[  1+\frac{\epsilon_k(\epsilon_k+\omega_0)}{4J^2} \right] \right.\\
&& +   \left. \frac{n_b+1-f(\epsilon_k-\omega_0)}{\sqrt{4t^2-(\epsilon_k- \omega_0)^2}}
     \left[  1+\frac{\epsilon_k(\epsilon_k-\omega_0)}{4J^2} \right] \right\rbrace, 
   \label{eq:rateuncorrqp}
\end{eqnarray}
\begin{eqnarray} \nonumber
  && \frac{1}{\tau_k} = 
    4g_\delta^2 \left\lbrace\frac{n_b+f(\epsilon_k+\omega_0)}{\sqrt{4J^2-(\epsilon_k+ \omega_0)^2}}
     \left[  2-\frac{\omega_0(\epsilon_k+\omega_0)}{4J^2} \right] \right.\\
&& +   \left.\frac{n_b+1-f(\epsilon_k-\omega_0)}{\sqrt{4t^2-(\epsilon_k- \omega_0)^2}}
     \left[  2+\frac{\omega_0(\epsilon_k-\omega_0)}{4J^2} \right] \right\rbrace,
   \label{eq:rateuncorr}
\end{eqnarray}
and for diagonal disorder:
\begin{eqnarray} \nonumber
   \frac{1}{\tau_k}  =\frac{1}{\tau_k^{qp}}  &=&
    2g_H^2 \left\lbrace\frac{n_b+f(\epsilon_k+\omega_0)}{\sqrt{4J^2-(\epsilon_k+ \omega_0)^2}}
     \right.\\
&& +   \left.\frac{n_b+1-f(\epsilon_k-\omega_0)}{\sqrt{4t^2-(\epsilon_k- \omega_0)^2}}
    \right\rbrace.
   \label{eq:rateHol}
\end{eqnarray}

\begin{figure}[h]
   \centering
      \includegraphics[width=8cm]{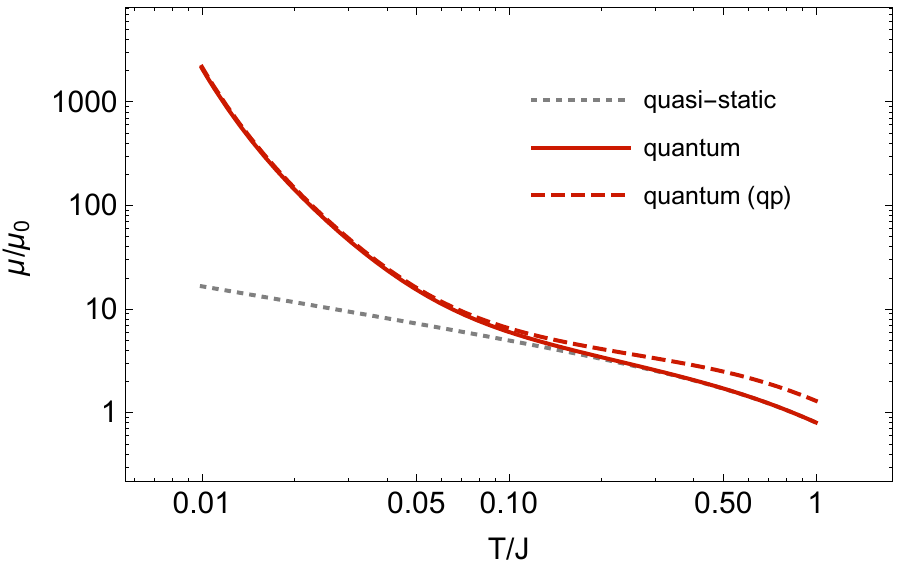}
      \includegraphics[width=8cm]{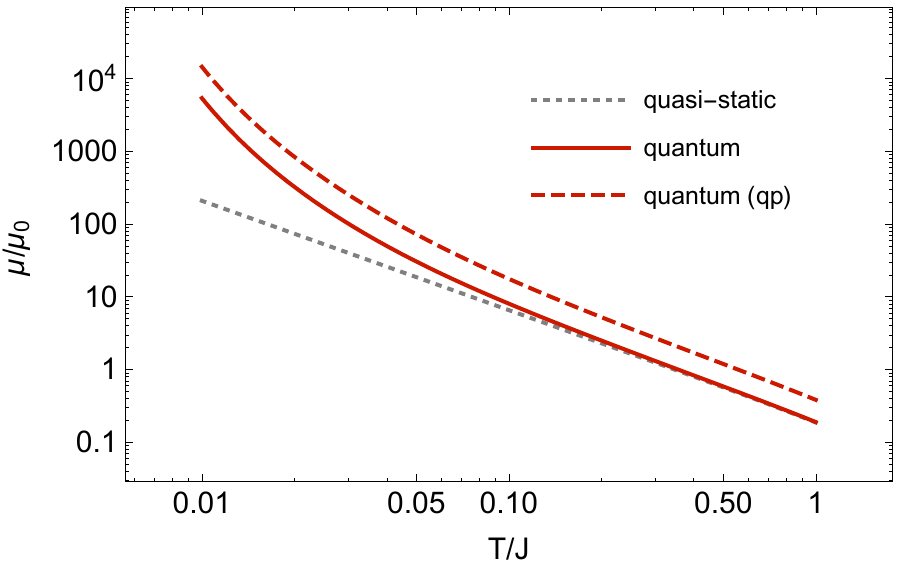}
  \caption{
   Quantum phonon effects, brought by restoring a finite $\omega_0/J=0.05$,  
  illustrated  in the case of uncorrelated
  bond disorder (top panel), with the same parameters  as in Fig. \ref{fig:muB}, and
    for correlated bond disorder (bottom panel). In this case,
  for comparison with the results of Ref. \cite{DeFilippis-PRL15} we have chosen 
  $\lambda_{SSH}=0.17$, corresponding to $\Delta J/J=0.41$ 
  at $T/J=0.25$ in the quasi-static case.}
 \label{fig:muBq}
 \end{figure}

Fig.   \ref{fig:muBq} shows the comparison of the quasi-static result Eq. (\ref{eq:tauuncorrvtx}) (orange)
with the full quantum expression Eq. (\ref{eq:rateuncorr}) (blue), in the case of uncorrelated bond disorder (top panel).
The inclusion of quantum phonons has two effects: the first is that the amount of intermolecular fluctuations
increases as compared to the classical value, 
due to the fact that $1+2n_b> 2T/\omega_0$, with the classical limit only being attained asymptotically 
for $T\gg \omega_0$. In particular, intermolecular fluctuations do not vanish at zero temperature 
as predicted by the classical formula but rather saturate to $(\Delta J/J)^2= \lambda_\delta \omega_0/J$.
This effect tends to increase the scattering rate, and hence to decrease the mobility with respect to the classical value.
The second effect is that, due to energy conservation, scattering by phonons is suppressed
within a shell of $\pm \omega_0$ around $\epsilon_k$, which leads instead to an increase in mobility. 
As shown in Fig. \ref{fig:muBq}(a), the latter always dominates and the mobility with quantum phonons is
larger than what is predicted in the quasi-static limit. The correction is around $2\%$ at 
$T= 0.25 J= 5 \omega_0$, validating the use of the quasi-static expression when investigating
room temperature mobilities.  The discrepancy between the quantum and quasi-static results however 
grows rapidly upon reducing the temperature, 
(it is about a factor of $2$ already at $T= \omega_0$) so that one should use the full quantum expression when addressing 
the low-temperature properties. The apparent power-law exponent $\gamma$ also rapidly increases upon reducing the temperature.
The bottom panel in Fig. \ref{fig:muBq} shows an analogous comparison  in the case of correlated bond disorder [Eq. (\ref{eq:taucorrvtx}) (orange),
Eq. (\ref{eq:rateSSH}) (blue)], showing a qualitatively similar behavior.
The correction in this case is respectively $7\%$ at $T=5 \omega_0$, and $60\%$ at $T= \omega_0$.

The results reported in Fig. \ref{fig:muBq} also indicate that the effect of the geometrical vertex
can be slightly modified by the inclusion of phonon quantum  fluctuations. This is especially true 
in the case of correlated disorder, where  
the exact reduction of a factor $2$ in the mobility in the quasi-static phonon limit 
[cf. Eqs. (\ref{eq:taucorrvtx})  and (\ref{eq:taucorr})], is increased 
in the quantum phonon case as soon as $T\lesssim \omega_0$, reaching a factor $3$ when $T\ll \omega_0$.
%provided that the phonon frequency $\omega_0\ll J$ (here $\omega_0/J=0.05$).
For uncorrelated disorder, vertex corrections become negligible at low temperature both in the quasi-static and in the quantum 
phonon case.

\subsection{Errata on previously reported results}
Eq. (\ref{eq:muBcorr1D}) for correlated bond disorder 
correctly appears in Refs. \cite{Fratini09} and \cite{Fratini-AFM16}. In both cases, however,
there is a misprint in the low-temperature expansion, which appears with the numerical prefactor $1/8$ 
instead of $1/16$, 
which  actually corresponds to the result calculated in the absence of the geometrical vertex $F_{k,k+q}$ 
(this result was used in the 
figures of these papers in order to provide a direct comparison with the results of the Kubo bubble apporximation,
which does not contain vertex corrections). 
An analogous overestimate of the mobility is reported 
in Fig. 1(d) of Ref. \cite{DeFilippis-PRL15}. The correct band theory result is the one reported in 
Fig. \ref{fig:validation}(a) above.

Similarly, the Boltzmann result reported as a dashed line in Fig. 3b of Ref. \cite{NMat17} for uncorrelated bond disorder 
was also overestimated by a factor of $2$. The correct result therefore moves closer to the transient 
localization value (see also Fig. 1b of the present work), 
although the significant discrepancies pointed out in that work remain. The temperature exponent  
shown as a dashed line in Fig. 3a of Ref. \cite{NMat17} is instead correct.

\section{Localization corrections and crossover to the transient localization regime}
\label{sec:crossover}
\subsection{Optical conductivity}
To evaluate the optical conductivity 
 we consider the Kubo formula
\begin{equation}
\sigma(\omega)=\frac{Re C_- (\omega)}{\omega}
\label{eq:Kubo}
\end{equation}
where $C_-(\omega)$ is the current-commutator correlation function (we  use units in which $e=\hbar=1$ and a 
unit lattice spacing).
The detailed balance condition relates $C_-(\omega)$ to the anticommutator correlation function $C(\omega)$
\begin{equation}
Re C_-(\omega)=\tanh(\beta \omega/2) Re C(\omega).
\label{eq:detailed}
\end{equation}
Combining Eqs. (\ref{eq:Kubo},\ref{eq:detailed}) yields
\begin{equation}
\sigma(\omega)=\frac{1}{\omega}\tanh(\beta \omega/2) Re C(\omega).
\end{equation}

From   Eq. (\ref{eq:CRTE}) of the main text 
we obtain
\begin{equation}
C(\omega) = {\cal FT} [C_{SC}(t) + (C_0(t)-C_{SC}(t))e^{-p t} ].
\label{eq:interpola}
\end{equation}
where ${\cal FT}$ indicates the Fourier transform and $p=\tau^{-1}_d$ is the inverse of a velocity decorrelation 
time (see text and Appendix \ref{sec:Thouless} below).
Substituting this expression  in the exact relations Eqs. (\ref{eq:Kubo},\ref{eq:detailed}) yields, 
for the real part of the optical conductivity:
\begin{equation}
\sigma(\omega)=\sigma_{SC}(\omega)+\sigma_{0}(\omega,p)-\sigma_{SC}(\omega,p)
\label{eq:splitsigma}
\end{equation}
where %the relaxation time approximation (RTA) term is
\begin{eqnarray}
\sigma_{0}(\omega,p)&=&\frac{1}{\omega}\tanh(\beta \omega/2) Re {\cal FT} [C_0(t)e^{-p t}]\label{eq:sigmaRTA}\\
\sigma_{SC}(\omega,p)&=&\frac{1}{\omega}\tanh(\beta \omega/2) Re {\cal FT} [C_{SC}(t)e^{-p t}]
\label{eq:sigmaSC}
\end{eqnarray}
and the semiclassical term in Eq. (\ref{eq:splitsigma}) is given by Eq. (\ref{eq:optavg}), i.e. 
\begin{equation}
\sigma_{SC}(\omega)=\frac{\sigma_{SC}(0)}{1+(\omega\tau)^2}
\label{eq:sigmaSC2}
\end{equation}
where $\sigma_{SC}(0)$ and $\tau$ are respectively the DC conductivity and the relaxation time obtained within 
the semiclassical approximation \cite{Allen} and evaluated in Section \ref{sec:band}.

The term in Eq. (\ref{eq:sigmaSC}) can be calculated via the
the Lehman representation (Ref. \cite{Ciuchi12}, Eq. (A12)) as
\begin{eqnarray}
\sigma_{SC}(\omega,p)& =& \frac{1}{\pi\omega}\tanh(\beta \omega/2)
\int_0^\infty d\nu \frac{\sigma_{SC}(\nu)\nu}{\tanh(\beta \nu/2)} \times \nonumber \\
& \times & \left [  \frac{p}{(\omega+\nu)^2+p^2}+\frac{p}{(\omega-\nu)^2+p^2} \right].
\label{eq:sigmaRTAp}
\end{eqnarray}
Direct inspection of Eq. (\ref{eq:sigmaRTAp}) shows that 
$\lim_{p\rightarrow 0} \sigma_{SC}(\omega,p)=\sigma_{SC}(\omega)$, so that 
Eq. (\ref{eq:splitsigma}) recovers the optical conductivity of the statically disordered Hamiltonian when $p\to 0$.
In the opposite  limit, $p\rightarrow \infty$,  the contributions from 
Eqs. (\ref{eq:sigmaRTA},\ref{eq:sigmaSC}) vanish as can  be verified explicitly from the form 
Eq. (\ref{eq:sigmaRTAp}). It is worth noting that at any finite $p$, $\sigma_{SC}(\omega,p)$ 
is not a simple Lorentzian convolution of $\sigma_{SC}(\omega)$.

\subsection{Mobility}
The quantum corrections to the mobility are explicitly derived in the main text. 
Here we show how these can be alternatively obtained as the $\omega\rightarrow 0$ limit of the AC conductivity.
Using the relation \cite{Ciuchi12}
\begin{equation}
\label{eq:CpCw}
C(p) = \int_0^\infty \frac{d\omega}{\pi} \frac{ 2p \omega \sigma(\omega)}{(p^2+\omega^2)\tanh (\beta \omega/2)}
\end{equation}
we obtain
\begin{equation}
\mu=\mu_{band}+\frac{\beta}{2} \left [ C_0(p)- C_{SC}(p)\right  ]
\label{eq:splitmu}
\end{equation}
which is  Eq. (\ref{eq:muRTE}) in the main text.

For practical calculations, we recognize that the second term in Eq. (\ref{eq:splitmu}) is the 
mobility $\mu_{TL}$ obtained from transient localization theory, which can be calculated by 
standard methods\cite{NMat17,Nemati}. To evaluate the remaining terms we
perform explicitly the integral in Eq. (\ref{eq:sigmaRTAp}) and take the limit $\omega\to 0$, which yields the  
final expression
\begin{eqnarray}
 \mu&=&\mu_{TL}+
\mu_{band} \left\lbrace \left\lbrack 1-\frac{\beta}{2\tau}\frac{\cot(\beta/2\tau)}{1+p\tau} 
\right\rbrack  +\frac{\frac{p\beta}{2\pi}}{1-(p\tau)^2} \times \right. \label{eq:mupractical}\\
&&\times \left.         
\left\lbrack 
2 h\left(\frac{p\beta}{2\pi}\right) +\left(1+\frac{1}{p\tau}\right)h\left(\frac{\beta}{2\pi \tau}\right)
+\left(1-\frac{1}{p\tau}\right)h\left(\frac{-\beta}{2\pi \tau}\right)
\right\rbrack
\right\rbrace,  \nonumber
\end{eqnarray}
where  $h(z)=\int_0^1 dx (1-x^z)/(1-x)$ denotes the integral representation  of the z$^{th}$ harmonic number. 
A similar expression can be obtained for the full frequency-dependent conductivity.

In the high temperature limit, $\beta p \ll 1$, Eq. (\ref{eq:mupractical}) expression simplifies to
\begin{eqnarray}
&& \mu=\mu_{TL}+
\frac{p\tau}{1+p\tau} \ \mu_{band}.
\label{eq:muhiT}
\end{eqnarray}

\subsection{Crossover condition}
\begin{figure}
   \centering
   \includegraphics[width=8cm]{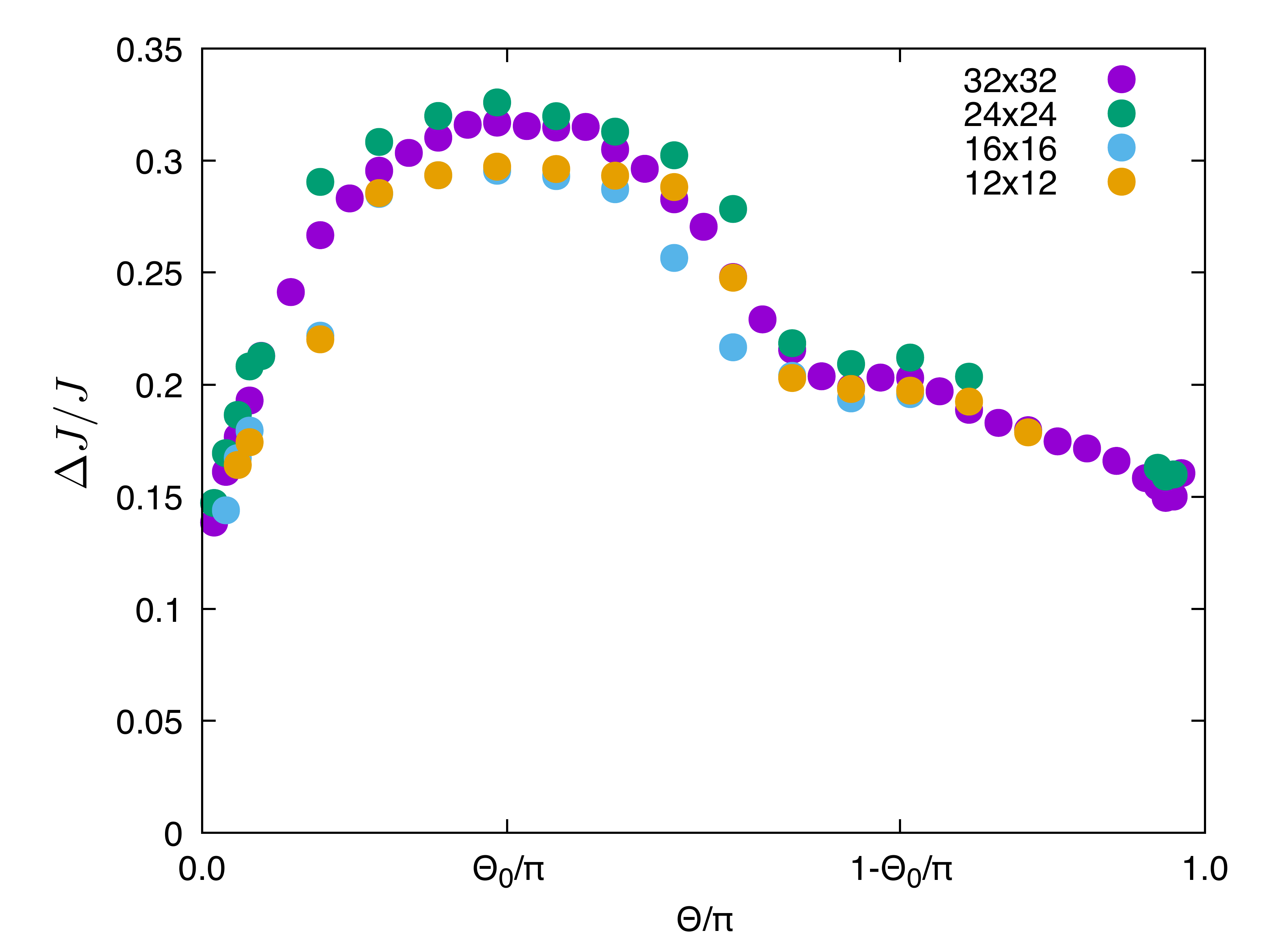}
 \caption{The crossover points obtained from 
 the condition $d^2\sigma/d\omega^2=0$, as a function of the band structure parameter $\theta$ defined in the main text, 
 for $T=0.25J$ and $1/\tau_d=0.05 J$. 
 Results are obtained for four different system sizes,  averaging
over 50 configurations of disorder.}
\label{fig:finitesize}
 \end{figure}

According to Eq. (\ref{eq:splitsigma}),  
the second derivative of the optical conductivity at $\omega=0$
consists of three parts. The semiclassical terms are directly evaluated from the explicit expression Eq. (\ref{eq:sigmaSC2}) 
\begin{eqnarray}
\frac{d^2\sigma_{SC}(\omega)}{d\omega^2}|_{\omega=0}&=&-2\sigma_{SC}(0)\tau^2
\label{eq:D2SC}\\
\frac{\partial^2\sigma_{SC}(\omega,p)}{\partial\omega^2}|_{\omega=0}&=&
-2\int_0^\infty \frac{d\nu}{\pi} \frac{\nu\sigma_{SC}(\nu)\phi(p,\nu)}{\tanh(\beta \nu/2)} 
\label{eq:D2SCp}
\end{eqnarray}
with
\begin{equation}
\phi(p,\nu)=\frac{\beta}{2}\frac{p}{p^2+\nu^2} \left [ -\frac{\beta}{6}+\frac{2(3\nu^2-p^2)}{(p^2+\nu^2)^2}\right ].
\end{equation}
The transient localization term is calculated from the exact
eigenvectors ($|n\rangle$) and eigenvalues ($E_n$) of the
statically disordered Hamiltonian as 
\begin{equation}
\frac{\partial^2\sigma_{0}(\omega,p)}{\partial\omega^2}|_{\omega=0}=
\frac{2}{Z}\sum_{n,m}e^{-\beta E_n}|\langle n|J|m \rangle|^2\phi(p,\omega_{nm})
\label{eq:D2RTA}
\end{equation}
where we have defined $\omega_{nm}=E_n-E_m$.
The crossover from band transport to transient localization is found when the sum of the three terms 
appearing in Eqs. (\ref{eq:D2SC},\ref{eq:D2SCp},\ref{eq:D2RTA}) 
vanishes. Fig. \ref{fig:finitesize} illustrates 
the crossover line obtained for $p=0.05$ ($\tau_d=20\hbar/J$) using four different system sizes, showing that 
finite-size effects are well controlled. 
The results reported in Fig. \ref{fig:PD} of the main text correspond to the largest  clusters studied, consisting of 
32x32 sites.

Similar to Eq. (\ref{eq:muhiT}), a compact analytical expression for the term appearing in Eq. (\ref{eq:D2SCp}) 
can be obtained in the large temperature limit $\beta p \ll 1$, 
which corresponds to
replacing Eq (\ref{eq:splitsigma}) with the simpler form 
\begin{equation}
\sigma(\omega)=\sigma_{SC}(\omega)+\sigma_{0}(\omega+ip)-\sigma_{SC}(\omega+ip)
\label{eq:splitsigmanapoli}
\end{equation}
with $\sigma_{0}(\omega)$ the optical conductivity of the the random static Hamiltonian. 
% (note that  $\sigma$ is taken here as a complex quantity.
Taking the second derivative yields
\begin{equation}
\frac{d^2\sigma(\omega)}{d\omega^2}|_{\omega=0}=\frac{\partial^2\sigma_{0}(\omega,p)}{\partial\omega^2}|_{\omega=0}+
2\sigma_{SC}(0)\tau^2 \frac{1-(1+p\tau)^3}{(1+p\tau)^3}.
\end{equation}
This approximate expression gives results comparable to the original Eq. (\ref{eq:splitsigma}) in all the cases 
studied in the main text.

\section{Thouless argument and identification of the parameter $\tau_d$}
\label{sec:Thouless}

The effects of decorrelation due to dynamic degrees of freedom in
disordered systems have been studied within the frame of the scaling theory of
localization in the past \cite{RMP84}. The typical situation addressed in
these works corresponds to disordered systems in which  a (small)
source of inelastic scattering is added, e.g. due to interaction of charge carriers with
phonons. 
Here we show how the Thouless argument \cite{Thouless}, that was devised to deal with this situation,
can be generalized to the case of purely dynamic disorder (i.e. where 
the lattice vibrations act  both as the source of disorder and of decorrelation) 
provided that a correct identification of the decorrelation time is made.

According to the Thouless argument, in a disordered system in the presence of inelastic scattering of phononic or electronic origin,
there exists a finite length scale $L_{Th}$  beyond which 
diffusive motion can take place. The  Thouless length is related to the  inelastic scattering time $\tau_{in}$ via 
$L^2_{Th}=D_{SC}\tau_{in}$, i.e. it can be interpreted as the diffusion length on the
inelastic scattering timescale, which acts as a decorrelation time. 
According to the scaling theory of localization in two dimensions, this corresponds to a finite DC conductivity \cite{RMP84} 
\begin{equation}
\sigma_{2D}(L_{th})=\sigma_{SC}+\delta \sigma_{DC}(L_{th})
\label{sigma2D}
\end{equation}
where the two-dimensional correction to the conductivity  due to weak localization is
\begin{equation}
\delta \sigma_{DC}(L_{th})=\frac{e^2}{2\pi^2\hbar} \log \frac{\ell^2}{L^2_{Th}},
\label{Deltasigma2D}
\end{equation}
with $\ell$ a microscopic length enforcing a lower cutoff in the scaling theory.
The presence of a finite length ($L^2_{Th}$) in the scaling term Eq. (\ref{sigma2D}) 
implies that the DC conductivity is non-vanishing, at variance with the case of a purely static disorder.
We can interpret this result in terms of an infrared cutoff $\omega^*$ in the quantum correction of the 
AC conductivity, by imposing
\begin{equation}
\delta \sigma_{DC}(L_{th})=\delta \sigma_{AC}(\omega^*).
\label{QC2d}
\end{equation}
To determine the corresponding $\omega^*$, we make use   of 
the scaling theory expression for the AC conductivity \cite{RMP84} 
\begin{equation}
\delta \sigma_{AC}(\omega)=\frac{\sigma_{SC}}{k_F\ell} \log |\omega\tau|.
\label{DeltasigmaAC}
\end{equation}
Equating the above result with Eq. (\ref{Deltasigma2D}) yields
\begin{equation}
\omega^*=\frac{\ell^2}{L^2_{Th}\tau}.
\label{omegain}
\end{equation}
If $\tau_{in}\gg \tau$,  the microscopic length $\ell$ can be identified with 
the semiclassical diffusion length, 
$\ell^2 \simeq D_{SC} \tau$, and consequently 
$\omega^*=1/\tau_{in}$.

Eq. (\ref{eq:splitsigmanapoli}) allows us to directly compare our results with that coming from the previous argument.
Let us now consider Eq. (\ref{eq:splitsigmanapoli}). In the $\omega=0$ limit, the 
quantum corrections to the  conductivity are
\begin{equation}
\delta \sigma=Re\sigma_0(ip)-Re\sigma_{SC}(ip).
\label{pippo}
\end{equation}
Continuing Eq. (\ref{DeltasigmaAC})   to the complex plane and substituting into Eq. (\ref{pippo}) yields 
\begin{equation}
\delta \sigma=\sigma_{SC}\left [  \frac{1}{k_F \ell} \log (p\tau) +\frac{p\tau}{p\tau-1}\right ] .
\label{eq:DeltaQC}
\end{equation}
Comparing Eq. (\ref{eq:DeltaQC}) in the  $p \tau \ll 1$ limit with Eq. (\ref{QC2d}) we get $p=\omega^*=1/\tau_{in}$. 
Since $p=1/\tau_d$,  we can therefore identify the inelastic time in the Thouless argument with the decorrelation time
introduced in the main text.

We note that in the standard situation where preexisting localization effects are decorrelated by
the inclusion of phonons, the decorrelation time is expected to diverge when the electron-phonon coupling strength
vanishes. In the case considered in this work, instead, where the 
disorder is itself of dynamical origin, the decorrelation time diverges when the
disorder becomes static, i.e when the {\em phonon frequency} $\omega_0$ vanishes. It
is then natural to assume \cite{Shante78,Gogolin87} $p\propto \omega_0$.
Beyond the analytical argument given here, 
the comparison with the exact QMC results shown in Fig. \ref{fig:validation}(a) for the correlated bond disorder model, 
and with the FTLM results of Fig. \ref{fig:validation}(b) for the Holstein model confirms the proprotionality between $p$ and 
$\omega_0$, and  indicates that the choice $p\simeq 2.2\omega_0 $ provides the most accurate quantitative results.
This identification is also compatible with the  values in the range $p=2-2.9\omega_0$ inferred from best fits of the Ehrenfest dynamics of 
Ref. \cite{Ciuchi11}, performed at various values of $\lambda$ and $\omega_0$ in the strong disorder regime.

\section{Details of numerical methods}
\label{sec:nummethods}
For the practical calculation of the mobility, of the instantaneous diffusivity and of the optical conductivity, 
we use the decomposition of the correlator given in Eq. (\ref{eq:Clap2}). The mobility, in particular, 
can be evaluated using the explicit form Eq. (\ref{eq:mupractical}), which is based on 
the single scattering time approximation Eq. (\ref{eq:tauavg})
as described in Appendix \ref{sec:models}. 

To evaluate the remaining TL part $\mu_{TL}$, one needs to solve the model Eq. (\ref{eq:H}) 
in the limit of static disorder, where the intermolecular transfer integrals form a statistical ensemble 
of gaussianly distributed variables with variances $\Delta J=\alpha_\delta\sqrt{k_BT/K}$. 
This is done numerically, using two different methods as documented in previous works.

\begin{itemize}
\item for the study of long time/long distance dynamics, needed in Fig. 1c and 1d, 
we implement the direct solution of the Schr\"odinger equation in the time domain
using the time diffusion method of Refs. \cite{Mayou00,NMat17}. This method allows to reach system sizes of
400x400 molecular sites, and times $t\sim 2000 \hbar/J$, at modest computational cost.

\item for all other cases, i.e. for the calculation of  
the quantum corrections to the mobility and for the determination of the crossover from 
band transport to transient localization, we use the exact diagonalization method 
described in Ref. \cite{Nemati}. The maximum system size studied here, of 32x32 sites, is sufficient to
reach convergence on the quantities of interest (see Fig. \ref{fig:finitesize}).
\end{itemize}

\begin{acknowledgments}
The authors are grateful to G. D'Avino and G. Schweicher for their valuable suggestions. 
S.F. acknowledges support by DFG (Grant No. DR228/48-1). 
\end{acknowledgments}

\end{document}